\documentclass{article}

\usepackage{graphicx}
\usepackage{cite}
\usepackage[margin=1in]{geometry}
\usepackage[dvipsnames]{xcolor}
\usepackage{comment} 
\usepackage{amsmath,amsthm,amssymb}
\usepackage{cancel}
\usepackage[ruled,vlined]{algorithm2e}
\usepackage{algorithmic}
\usepackage[hyphens]{url}

\usepackage{caption}
\usepackage{subcaption}
\usepackage{enumitem}
\usepackage[normalem]{ulem}
\usepackage{bm}

\usepackage{hyperref}

\usepackage{authblk}

\newtheorem{theorem}{Theorem}[section]
\newtheorem{lemma}[theorem]{Lemma}
\newtheorem{proposition}[theorem]{Proposition}

\theoremstyle{definition}
\newtheorem{definition}[theorem]{Definition}

\theoremstyle{remark}
\newtheorem{remark}[theorem]{Remark}

\numberwithin{equation}{section}

\DeclareMathOperator*{\argmin}{argmin}

\newcommand{\numreviewers}{n}
\newcommand{\numpapers}{m}
\newcommand{\reviewperpaper}{k}
\newcommand{\paperperreviewer}{\ell}

\newcommand{\reviewerSet}{[\numreviewers]}
\newcommand{\paperSet}{[\numpapers]}

\newcommand{\resultscorevec}{(\theta^*_1, \theta^*_2, ..., \theta^*_\numreviewers)}
\newcommand{\resultscoresym}{\bm{\theta}^*}
\newcommand{\resultscore}{\theta^*}
\newcommand{\resultscoresymproj}{{{\bm{\theta}}}}

\newcommand{\resultscoreproj}{\theta}
\newcommand{\paperperreviewerset}{\mathcal{L}}

\newcommand{\releasescorevec}{(r_1, r_2, ..., r_\numreviewers)}
\newcommand{\releasescoresym}{\textbf{r}}
\newcommand{\releasescore}{r}

\newcommand{\noisescorevec}{(\eta_1, \eta_2, ..., \eta_\numreviewers)}
\newcommand{\noisescoresym}{\bm{\eta}}
\newcommand{\noisescore}{\eta}

\newcommand{\finalscoresym}{\bm{t}}
\newcommand{\finalscore}{t}

\newcommand{\samplelist}{\mathcal{W}}
\newcommand{\samplelistitem}{w}
\newcommand{\resultscorefirst}{w_1}
\newcommand{\resultscoresecond}{w_2}
\newcommand{\samplelistsize}{n}
\newcommand{\samplelistavg}{a}

\newcommand{\sampleindex}{i}
\newcommand{\sampleindexscd}{j}
\newcommand{\sampleindexscdnew}{j'}

\newcommand{\receivescore}{x}
\newcommand{\samplescore}{z}

\newcommand{\totalgivenscore}{y}
\newcommand{\totalgivenscorevec}{(y_1, y_2, ..., y_\numreviewers)}
\newcommand{\reviewPaperSet}{\mathcal{Y}}

\newcommand{\rawscore}{s}
\newcommand{\normalizedsub}{\widetilde{\rawscore}}
\newcommand{\paperreviewerscore}{w}

\newcommand{\assignmentset}{\mathcal{G}}
\newcommand{\samplegraph}{\textit{g}}

\newcommand{\scoreset}{\bm{\Theta}}
\newcommand{\convexSet}{\mathcal{C}}

\newcommand{\scorematrix}{X}

\newcommand{\partitionset}{\mathcal{V}}

\newcommand{\firstpartitionset}{\mathcal{S}_1}
\newcommand{\restpartitionset}{\mathcal{S}}

\newcommand{\noisydata}{v}

\newcommand{\alltuples}{\Omega'}
\newcommand{\sortedalltuples}{\Omega}
\newcommand{\tuplegraph}{G}
\newcommand{\tuplevertex}{\nu}

\newcommand{\Lo}{L}
\newcommand{\Up}{U}

\begin{document}
\title{On the Privacy-Utility Tradeoff in Peer-Review Data Analysis}

\author{Wenxin Ding$^1$, Nihar B. Shah$^{1,2}$, Weina Wang$^1$ \\
Computer Science Department$^1$, Machine Learning Department$^2$\\
Carnegie Mellon University\\
\texttt{\{wenxind@andrew, nihars@cs, weinaw@cs\}.cmu.edu} \\
}

\date{}

\maketitle

\begin{abstract}
A major impediment to research on improving peer review is the unavailability of peer-review data, since any release of such data must grapple with the sensitivity of the peer review data in terms of protecting identities of reviewers from authors. We posit the need to develop techniques to release peer-review data in a privacy-preserving manner. Identifying this problem, in this paper we propose a framework for privacy-preserving release of certain conference peer-review data --- distributions of ratings, miscalibration, and subjectivity --- with an emphasis on the accuracy (or utility) of the released data. The crux of the framework lies in recognizing that a part of the data pertaining to the reviews is already available in public, and we use this information to post-process the data released by any privacy mechanism in a manner that improves the accuracy (utility) of the data while retaining the privacy guarantees. Our framework works with any privacy-preserving mechanism that operates via releasing perturbed data. We present several positive and negative theoretical results, including a polynomial-time algorithm for improving on the privacy-utility tradeoff.
\end{abstract}

\section{Introduction}
\label{sec:intro}

A fair and efficient peer-review process is of utmost importance to the development of scientific research. There are, however, a large number of challenges in peer review, pertaining to its fairness and efficiency. Consequently there is an overwhelming desire to ``fix'' the ``broken'' peer review process~\cite{rennie2016make,mccook2006peer}. And taking heed to this call, there is a growing amount of research on this topic. 

Research on improving peer review suffers from a considerable handicap -- unavailability of data~\cite{balietti2016science,tomkins2017reviewer,squazzoni2020unlock,schroter2020research}. Concealing the identities of reviewers from authors of any paper is paramount in most peer review systems. Thus releasing any peer review data is fraught with the risk of compromising on this privacy. 

As noted by Balietti et al.~\cite{balietti2016science}:
\begin{quote}
\emph{``The main reason behind the lack of empirical studies on peer-review is the difficulty in accessing data. In fact, peer-review data is considered very sensitive, and it is very seldom released for scrutiny, even in an anonymous form.''}
\end{quote}

Although there is a large body of research on the topic of privacy in various domains, not much privacy research directly targets the application of peer review. In an influential recent paper~\cite{tomkins2017reviewer} about peer review, Tomkins, Zhang and Heavlin highlight the challenges they faced in this respect and their consequent inability to release data:
\begin{quote}
\emph{``We would prefer to make available the raw data used in our study, but after some effort we have not been able to devise an anonymization scheme that will simultaneously protect the identities of the parties involved and allow accurate aggregate statistical analysis. We are familiar with the literature around privacy preserving dissemination of data for statistical analysis and feel that releasing our data is not possible using current state-of-the-art techniques.''}
\end{quote} 

We thus posit the need to develop techniques to help release peer-review data while ensuring that identities of reviewers of any paper are protected. With this motivation, we focus on the privacy-utility tradeoff in releasing certain conference peer-review data. The data to be released comprises distributions of the ratings or miscalibration or subjectivity in the peer-review process. The notion of privacy we consider is quite general -- our techniques apply to any notion of privacy which operates by perturbing the data, including differential privacy. We design a framework to improve in this tradeoff by improving the utility (accuracy) while retaining privacy guarantees.

Our work relies on the key the observation that a non-trivial part of conference peer-review data is already available in the public domain. We design techniques which use this publicly available information to post-process the data released by any privacy mechanism. Our approach is guided by the following four desiderata for such a post processing:
\begin{enumerate}[label=D\arabic*]
    \item \label{itmMaintext:desideratum-accuracy}Under no circumstances should the accuracy go down after applying the algorithm.
    \item \label{itmMaintext:desideratum-privacy}Under no circumstances should the privacy guarantee be compromised after applying the algorithm.
    \item \label{itmMaintext:desideratum-time}The algorithm should have a computational complexity that is polynomial in the number of reviewers and papers.\footnote{In typical conferences, the number of papers per reviewer and the number of reviewers per paper are both constants~\cite{shah2018design}.}
    \item \label{itmMaintext:desideratum-axiomatic}In special cases where an exact answer can be easily obtained from public data, the algorithm should also return the same answer with no error. (This is defined formally in Section~\ref{sec:axioms}.)
  \end{enumerate}

Our technical contributions towards this problem are as follows. We first argue that projecting the (noisy) output of the privacy mechanism on  the convex hull of all possible true values is desirable from the perspective of the desiderata. We show that, however, such a projection is NP-hard (via reducing the $\paperperreviewer$-partition problem). We then design a polynomial-time computable algorithm which projects the noisy output of the privacy mechanism on a convex set containing all possible true values, and satisfies the four desiderata listed above. As a result of independent interest, we also prove that the more obvious approach of projecting on the set of all true values (instead of a convex set containing them) can, in fact, reduce the accuracy. Finally, we conduct synthetic simulations, which reveal that our methods can yield considerable improvements in the privacy-utility tradeoff as compared to standard approaches. The code for our algorithm is available here: 
\url{https://github.com/wenxind/privacy-utility-tradeoff-in-peer-review-data}.

\section{Related Work}

This work falls in the intersection of two lines of research: peer review and privacy.

\noindent{\bf Peer review:} Peer review is the backbone of scientific research. There is an overwhelming desire in many domains of science and engineering for improving peer review, and consequently, there are many past works on the topic of either evaluating the efficacy of peer review or improving the peer review process~\cite{peters1982peer-review,kliewer2004peer,bennett2018radiation,mavrogenis2020good,bernard2018gender,snodgrass06literature,scott1974interreferee,lindsey1988assessing,douceur2009paper,reinhart2009peer}. These works, however, largely focus on the journal reviewing setup that is common in non-computer-science fields, whereas our focus is on the conference reviewing setting which is more common in computer science. 

The number of submissions to many computer science conferences, particularly to machine learning or artificial intelligence conferences, is growing near-exponentially and is presently in the several thousands. This rapid growth has spurred a considerable amount of recent research on peer review in computer science. These works include those on handling problems related to reviewer-assignment~\cite{goldsmith07aiconf,charlin13tpms,welch2014referee,stelmakh2018forall,kobren19localfairness}, miscalibration~\cite{roos2011calibrate,ge13bias,wang2018your}, subjectivity~\cite{noothigattu2018choosing}, biases~\cite{tomkins2017reviewer,stelmakh2019testing}, strategic behavior~\cite{balietti2016science,xu2018strategyproof,stelmakh2020catch} and others~\cite{cabanac2013capitalizing,fiez2019super,nips14experiment,shah2018design,stelmakh2020resubmissions}. In particular, as will be detailed later, our work is also useful towards releasing data pertaining to miscalibration and subjectivity, thereby helping in the understanding and mitigation of these problems. ~\\

%Multiple research has been done in finding a more fair and efficient way to choose papers from peer-review. An approach was proposed by Noothigattu, Shah and Procaccia \cite{noothigattu2018choosing}. It identifies that the disparate mapping of criteria scores to final recommendations by different reviewers is a major source of inconsistency. However, this cannot conceal the fact that many research on peer-review is limited due to lack of released data. 

\noindent{\bf Privacy:}
Privacy-preserving data analytics has been receiving rapidly increasing attention as the big-data regime emerges. There is a large body of research that investigates formal notion of privacy and quantifies the tradeoff between privacy and utility (see, e.g., \cite{DwoMcSNis_06,Dwo_06,BluLigRot_08,GabAriHsu_14,WanYinZha_16_2,BunUllVad_18}). Among these studies, differential privacy \cite{DwoMcSNis_06,Dwo_06} has become the de facto standard and has been applied to many areas.

Relaxations of differential privacy have also been proposed to enable more accurate data analysis \cite{DwoKenMcS_06,DwoRot_16,BunSte_16,Mir_17,BunDwoRot_18}.
For example, consider application scenarios such as federated learning where an individual’s data may spread across multiple datasets. In such a scenario, the cumulative privacy loss of an individual needs to be constrained when each dataset releases its statistics with privacy-preserving mechanisms. The relaxation of $(\epsilon, \delta)$-differential privacy \cite{DwoKenMcS_06} allows the amplitude of the noise injected to each dataset to scale approximately as $\sqrt{k}$ with $k$ being the number of datasets that contain a single individual’s data. Such composition scalings can be further improved when the tail behavior of errors is of more interest \cite{BunDwoRot_18}.

In this paper, we investigate the privacy-utility tradeoff for publishing histograms of peer-review data. Privacy-preserving release of histograms has been a major focus of the literature~\cite{ChaDwoMcS_05,DwoKenMcS_06,hay2010boosting,LiHayRas_10,BasSmi_15,BalVad_19}. 
To the best of our knowledge, existing techniques for improving the privacy-utility tradeoff are generally inadequate for the application of peer review since they do not take into account the special structures in peer-review data. For example, as we pointed out, one special feature of peer-review data is that part of the data is already publicly available in a non-privacy-preserving form. As we discuss in the sequel, the idiosyncratic nature of the peer-review setting implies that one can design methods tailored to this application which yield a (considerable) improvement in the privacy-utility tradeoff as compared to standard privacy mechanisms.
We comment that the techniques we develop take inspiration from the constrained inference technique in \cite{hay2010boosting}, which enforces consistency among the noisy answers to multiple queries. But again, application scenarios of the approach in \cite{hay2010boosting} do not possess the distinctive structures of peer-review data.
We reiterate that our approach of improving accuracy is not specific to a particular privacy notion, but rather it is a post-processing framework that applies to any privacy-preserving mechanism. Such mechanisms include the widely used Laplace mechanism (which guarantees differential privacy), Gaussian mechanism \cite{DwoKenMcS_06,DwoRot_16,Mir_17}, Sinh-Normal mechanism \cite{BunDwoRot_18}, as well as mechanisms that do not operate via addition of noise.~\\

\noindent{\bf Peer review and privacy:}
An exception is the concurrent work~\cite{jecmen2020manipulation} which considers releasing the reviewer-paper similarity matrix and source code for the reviewer assignment (whereas in contrast we consider releasing a function of the scores given by reviewers to papers). Their approach involves modifying and randomizing the reviewer-paper assignment process and their guarantees pertain to plausible deniability (that is, any reviewer may be assigned to any paper with a probability at most a certain value). On the other hand, we do not modify the peer-review process in any way, and instead use any privacy-preserving data-release mechanism coupled with post processing of the data from peer review.

\section{Background and problem setting}
In this section, we provide some background on the peer review setting and privacy, and describe our problem setting in more detail.

\subsection{Peer review}

We consider a conference peer review setting, where there are $\numreviewers$ reviewers and $\numpapers$ papers. We index the papers as $[\numpapers]=\{1,2, \cdots, \numpapers\}$ and the reviewers as $[\numreviewers]=\{1,2, \cdots, \numreviewers\}$.\footnote{We follow the standard convention of using $[\beta]$ to represent the set $\{1,2,\ldots,\beta\}$ for any positive integer $\kappa$.} For simplicity we assume that the number of papers reviewed by each reviewer is the same for all reviewers -- denoted as $\paperperreviewer$, and that the number of reviewers reviewing each paper is the same for all papers -- denoted as $\reviewperpaper$.\footnote{Our work is also applicable to the most general setting in which different reviewers and/or different papers have different loads. We discuss this in Section~\ref{sec:appendix:fullalgo}.} Consequently, we have the relation $\numreviewers  \paperperreviewer = \numpapers  \reviewperpaper$. All four parameters $(\numreviewers, \numpapers, \paperperreviewer, \reviewperpaper)$ are public knowledge.

Each review comprises a real-valued score. We assume that all papers and all associated reviews (that is, the set of scores received by each paper) are public knowledge (e.g., in conferences such as ICLR and others on the OpenReview.net review platform). The list of all reviewers is also available publicly (such a list is released by many conferences). However, importantly, the identity of which reviewer reviewed which paper is private.

We now introduce notation to describe the score given in any review. If reviewer $\sampleindexscd \in \reviewerSet$ reviews paper $\sampleindex \in \paperSet$, then we use $\rawscore_{\sampleindex \sampleindexscd} \in \mathbb{R}$ to denote the score of this review. This score is private in the sense that the identity of the reviewer who gives this score is not publicly available. However, for each paper $\sampleindex \in \paperSet$, the multiset $\{\rawscore_{\sampleindex \sampleindexscd} | \text{ reviewer } \sampleindexscd \in \reviewerSet \text{ reviews paper } \sampleindex\}$ is public.

\begin{figure}[t!]
\centering
\begin{subfigure}[b]{0.45\textwidth}
  \centering
  \includegraphics[width=1.1\columnwidth]{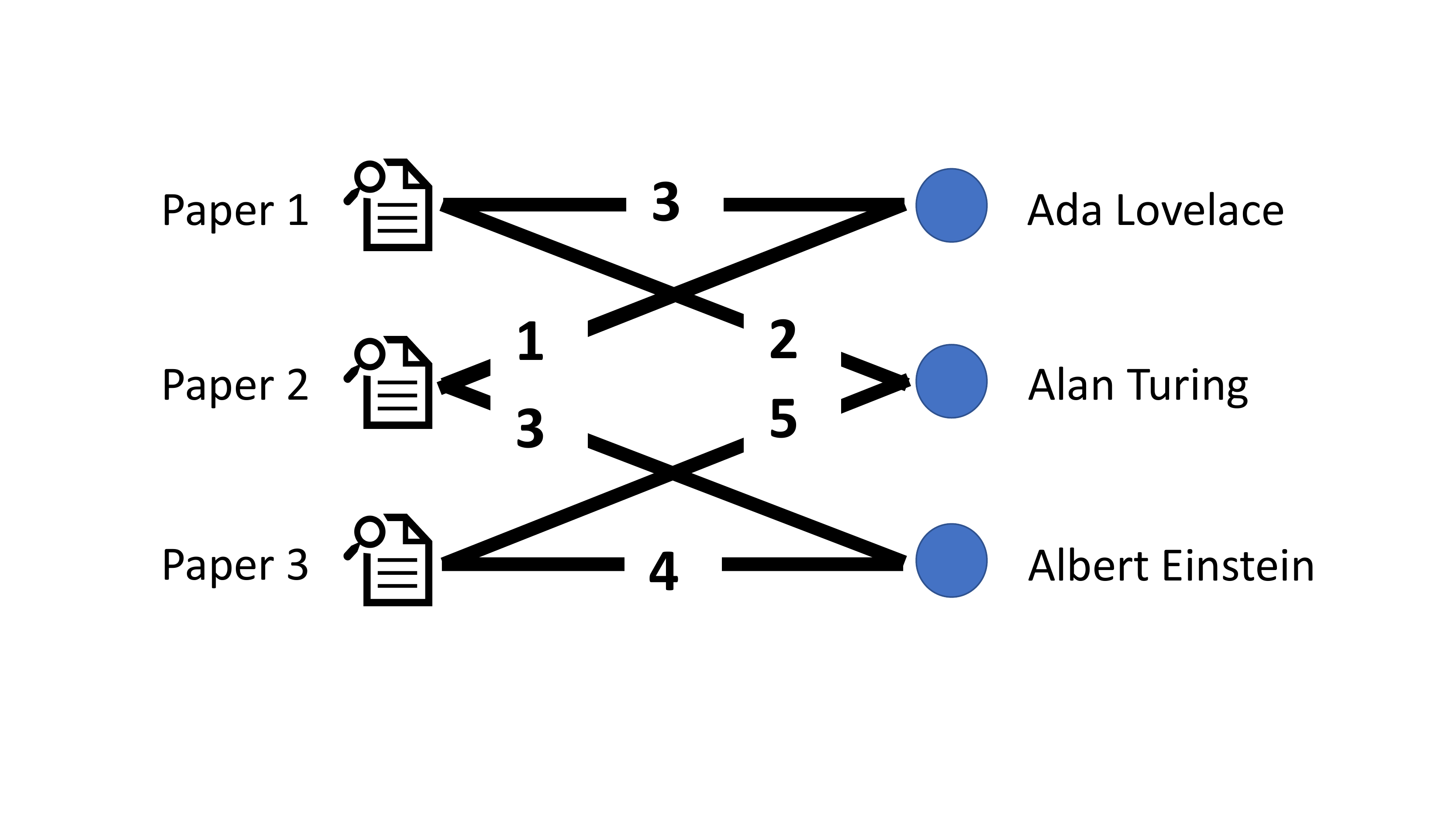}
  \caption{\label{fig:private_data}Private data}
\end{subfigure}\hspace{.09\textwidth}
\begin{subfigure}[b]{0.45\textwidth}
  \centering
  \includegraphics[width=1.1\columnwidth]{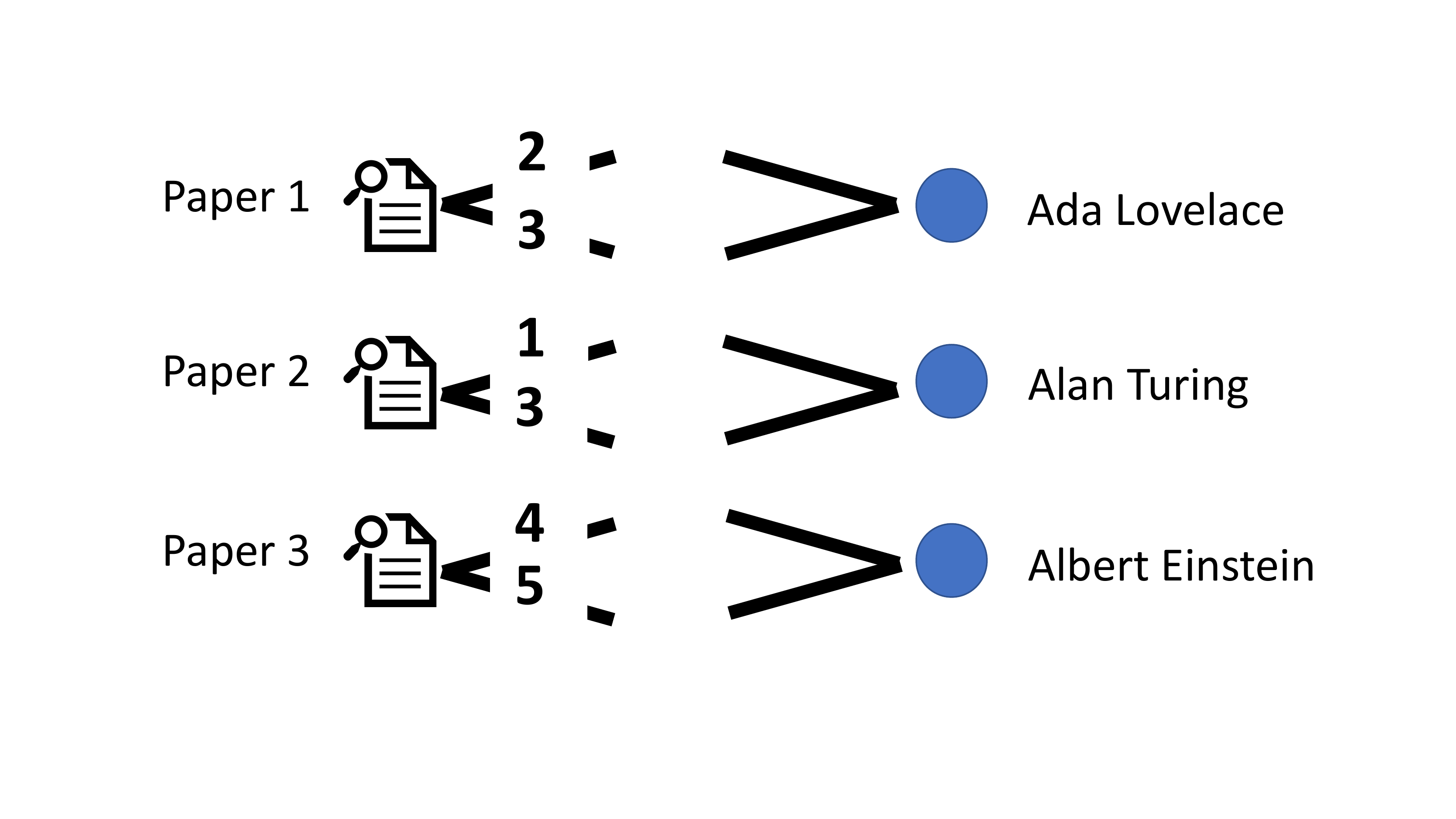}
  \caption{\label{fig:public_data}Public data} 
\end{subfigure}
\caption{\label{fig:bipartite}An illustration of the data (a) available privately to the program chairs of the conference, and (b) available to the public under increasingly popular `open review' paradigms in computer science.} 
\end{figure}

This setting can be described by a bipartite graph, as shown in Figure~\ref{fig:bipartite}. The bipartite graph has two disjoint sets of vertices, $\paperSet$ and $\reviewerSet$ representing the sets of papers and reviewers, repectively. In private data (Figure~\ref{fig:private_data}), an edge exists between any vertex (paper) $\sampleindex \in \paperSet$ and any vertex (reviewer) $\sampleindexscd \in \reviewerSet$ if reviewer $\sampleindexscd \in \reviewerSet$ reviews paper $\sampleindex \in \paperSet$. We associate each edge $(\sampleindex, \sampleindexscd)$ with the score $\rawscore_{\sampleindex \sampleindexscd}$. The edges (and their values) are all private. The private data is accessible to the program chairs of the conference. 
In public data (Figure~\ref{fig:public_data}), for each vertex (paper) in $\paperSet$, the weights of the edges connected to it are known publicly. However, the edges of the graph are not known. Note that in both public and private data, identities of papers and reviewers are known.

There are various quantities of interest for release that we consider in this work. An intermediate set of terms towards these quantities is the multiset $\{\paperreviewerscore_{\sampleindex \sampleindexscd} | \text{ reviewer } \sampleindexscd \in \reviewerSet \text{ reviews paper } \sampleindex \in \paperSet\}$ discussed below, which we refer to as the set of weights. This multiset can be computed from the scores $\{\rawscore_{\sampleindex \sampleindexscd} | \text{ reviewer } \sampleindexscd \in \reviewerSet \text{ reviews paper } \sampleindex \in \paperSet\}$.
We now discuss three such choices of $\{\paperreviewerscore_{\sampleindex \sampleindexscd} | \text{ reviewer } \sampleindexscd \in \reviewerSet \text{ reviews paper } \sampleindex \in \paperSet\}$, and subsequently describe the data we aim to release.

\begin{itemize}
\item \textbf{Reviewer ratings.} In this case, the mapping from scores to weights is simply the identity mapping:
\begin{equation}
   \paperreviewerscore_{\sampleindex \sampleindexscd} = \rawscore_{\sampleindex \sampleindexscd}.
   \label{eq:revratings}
\end{equation}

\item \textbf{Miscalibration.} Miscalibration is the problem that some reviewers are strict and some are lenient~\cite{roos2011calibrate,ge13bias,wang2018your}. In order to understand the amount of miscalibration, it is instructive to see the difference between the scores given by a reviewer and the scores given by other reviewers for the same papers. To this end, we let $\paperreviewerscore_{\sampleindex \sampleindexscd}$ denote the miscalibration in any individual review (for any paper $\sampleindex$ by any reviewer $\sampleindexscd$):
\begin{equation}
     \paperreviewerscore_{\sampleindex \sampleindexscd} = \rawscore_{\sampleindex \sampleindexscd} - \frac{1}{\reviewperpaper-1} \sum\limits_{\sampleindexscdnew \neq \sampleindexscd} \rawscore_{\sampleindex \sampleindexscdnew}.
\end{equation}

\item \textbf{Subjectivity.} Subjectivity is the problem that different reviewers place different emphasis on the various criteria when making an overall decision for a paper~\cite{lee2015commensuration}. Techniques such as that proposed in~\cite{noothigattu2018choosing} can be used to normalize each score in a manner that mitigates the subjectivity. Specifically, the technique in~\cite{noothigattu2018choosing} uses the public data to transform the score $\rawscore_{\sampleindex \sampleindexscd}$ associated to each review into a normalized version, say, $\normalizedsub_{\sampleindex \sampleindexscd}$. We can then set 
%the weights of the aforementioned graph as 
$\paperreviewerscore_{\sampleindex \sampleindexscd} = \normalizedsub_{\sampleindex \sampleindexscd}$ 
%between every paper $\sampleindex$ and reviewer $\sampleindexscd$
for every review, and the algorithm in this paper will help release statistics of these normalized scores. A second use case we consider is to better understand and investigate the issue of subjectivity, by releasing the amount of subjectivity present in the system, that is, the aggregate difference of reviewers' scores and their normalized scores. Concretely in this case, after obtaining the normalized scores 
$\{\normalizedsub_{\sampleindex \sampleindexscd} | \text{ reviewer } \sampleindexscd \in \reviewerSet \text{ reviews paper } \sampleindex \in \paperSet\}$, we set 
%the weights of the graph as
$\paperreviewerscore_{\sampleindex \sampleindexscd} = \rawscore_{\sampleindex \sampleindexscd} - \normalizedsub_{\sampleindex \sampleindexscd}$ for every review. %Again, note that in both the above cases, the transformation from $\rawscore$ to $\normalizedsub$ and $\paperreviewerscore$ can be done entirely using public data alone.
\end{itemize} 

Analogous to the scores $\rawscore_{\sampleindex \sampleindexscd}$'s, the weights $\paperreviewerscore_{\sampleindex \sampleindexscd}$'s are also associated to public and private components. Note that we can use the same bipartite graphs as in Figure~\ref{fig:bipartite} to represent the setting with weights. In particular, the private data continues to include the edges pertaining to which reviewer reviewed which paper. The private data also includes the weight $\paperreviewerscore_{\sampleindex \sampleindexscd}$ on each edge $(\sampleindex, \sampleindexscd)$ representing the weight that reviewer $\sampleindex \in \reviewerSet$ gives to paper $\sampleindexscd \in \paperSet$. The private data is depicted in Figure~\ref{fig:private_data} where in this interpretation, the values on the edges represent the weights. The public data, as in the case of scores, only includes the multiset of weights received by any paper, that is, the public data comprises the multisets $\{\paperreviewerscore_{\sampleindex \sampleindexscd} | \text{reviewer } \sampleindexscd \in \reviewerSet \text{ reviews paper } \sampleindex\}$ for every paper $\sampleindex \in \paperSet$. The public data is depicted in Figure~\ref{fig:public_data} where the values on the edges represent the weights.

It is very important to note the following two properties in the transformation of scores $\rawscore_{\sampleindex \sampleindexscd}$ to weights $\paperreviewerscore_{\sampleindex \sampleindexscd}$ for each of the aforementioned choices. First, clearly, given access to all private scores, all weights can be computed. Second, the public weights (that is, the multisets $\{\paperreviewerscore_{\sampleindex \sampleindexscd} | \text{reviewer } \sampleindexscd \in \reviewerSet \text{ reviews paper } \sampleindex\}$ for every paper $\sampleindex \in \paperSet$) can be computed using only the publicly available score data (that is, the multisets $\{\rawscore_{\sampleindex \sampleindexscd} | \text{reviewer } \sampleindexscd \in \reviewerSet \text{ reviews paper } \sampleindex\}$ for every paper $\sampleindex \in \paperSet$). This relation between the public (respectively, private) weights and public (respectively, private) scores allows us to interchange them in the graphs in Figure~\ref{fig:bipartite}. For simplicity, the reader may choose to simply consider the ``reviewer ratings'' choice~\eqref{eq:revratings} and think of simply the scores as the corresponding weights.

~\\ For each reviewer $\sampleindexscd \in \reviewerSet$, let $\reviewPaperSet_\sampleindexscd$ be the set of all papers reviewed by reviewer $\sampleindexscd$, that is, $\reviewPaperSet_\sampleindexscd = \{\sampleindex \in \paperSet \ | \ \text{reviewer } \sampleindexscd \text{ reviews paper } \sampleindex\}$. Let $\totalgivenscore_\sampleindexscd$ denote the mean weight of reviewer $\sampleindexscd$:
\begin{equation}
   \totalgivenscore_\sampleindexscd = \frac{1}{\paperperreviewer} \sum\limits_{\sampleindex \in \reviewPaperSet_\sampleindexscd} \paperreviewerscore_{\sampleindex \sampleindexscd}.
\end{equation}
Note that since the identity of the reviewer in any review is private, the values of $\reviewPaperSet_\sampleindexscd$ and $\totalgivenscore_\sampleindexscd$ in general cannot be computed from the public data.\\

{\bf Quantity to be released: }The quantity of interest is the histogram of the mean weights per reviewer, represented by the \emph{sorted} version of the mean-weight vector $\totalgivenscorevec$, which we denote by $\resultscoresym = \resultscorevec$. Then $\resultscore_1 \le ... \le \resultscore_\numreviewers$ and the multiset $\resultscorevec$ equals the multiset $\totalgivenscorevec$. We call $\resultscoresym$ the true sorted mean-weight vector. According to the applications discussed above, the vector $\resultscoresym$ can either represent the mean scores per reviewer or capture the amount of miscalibration, or subjectivity in the reviews. ~\\

Our goal is to release $\resultscoresym$, while ensuring privacy of reviewer identities. When the underlying weights are equal to the scores, the sorted mean-weight vector (that is, the histogram of scores) is commonly released by various conferences~\cite{shah2018design}. These are however usually released without any privacy considerations, and our work addresses privacy-preserving release with high accuracy. Addressing the issues of miscalibration and subjectivity is extremely important for fair and high-quality peer review~\cite{ge13bias,wang2018your,roos2011calibrate,lee2015commensuration,noothigattu2018choosing,siegelman1991assassins,kerr1977manuscript}, and releasing the statistics pertaining to the amount of miscaliabraiton or subjectivity can considerably help both research and policy-design regarding these issues.

Publishing histograms of datasets in a privacy-preserving manner has been a central objective in the literature of privacy research~\cite{ChaDwoMcS_05,DwoKenMcS_06,hay2010boosting,LiHayRas_10,BasSmi_15,BalVad_19}. Histograms are typically the most frequently used statistics in official reports, and more importantly, they form the basis for more complicated statistical analysis. To the best of our knowledge, existing techniques for improving the privacy-utility tradeoff (for histograms or otherwise) are generally inadequate for the application of peer review since they do not take into account the special structures in peer-review data. Our goal is to use the specific type of publicly available data in peer review in order to improve the privacy-utility tradeoff.

Finally we note that the high-level ideas behind our proposed algorithm are more general and may also be used to improve the utility of the released data for settings beyond the sorted mean-weight vector. We revisit this point later in the paper.

\subsection{Privacy}

To protect the privacy of reviewers, we consider privacy-preserving mechanisms that (randomly) perturb the quantities of interest. By virtue of the random perturbation, it is then hard to infer each individual reviewer's scores given to papers from the noisy data. 
Specifically, we consider any privacy mechanism that releases a vector $\releasescoresym = \releasescorevec \in \mathbb{R}^{\numreviewers}$ obtained possibly by adding random noise to the true sorted mean-weight vector $\resultscorevec \in \mathbb{R}^{\numreviewers}$. An example of such a privacy mechanism is the Laplace mechanism, which satisfies the popular notion of differential privacy~\cite{dwork2016calibrating,DwoRot_14} in which $
\releasescorevec = \resultscorevec + \noisescorevec$. Here $\noisescore_1,\noisescore_2,\ldots,\noisescore_n$ are i.i.d.\ random variables drawn from a zero-mean Laplace distribution.

\subsection{Utility (Accuracy)}

Let $\finalscoresym=(\finalscore_1, \cdots, \finalscore_\numreviewers)$ denote the final output (after post-processing) that is released. We measure the utility or accuracy of the output in terms of its \emph{mean squared error} with respect to the true value of the vector $\resultscoresym$, that is, $\mathbb{E} \left[\sum\limits_{\sampleindex = 1}^\numreviewers (\resultscore_\sampleindex - \finalscore_\sampleindex)^2 \right]$. We say that an (possibly random) output $\finalscoresym = (\finalscore_1, ..., \finalscore_\numreviewers)$ is more accurate than another output $\finalscoresym' = (\finalscore'_1, ..., \finalscore'_\numreviewers)$ with respect to $\resultscoresym$ if 
\begin{equation}
    \mathbb{E} \left[\sum\limits_{\sampleindex = 1}^\numreviewers (\resultscore_\sampleindex - \finalscore_\sampleindex)^2 \right] < \mathbb{E} \left[\sum\limits_{\sampleindex = 1}^\numreviewers (\resultscore_\sampleindex - \finalscore'_\sampleindex)^2 \right].
\end{equation}

\subsection{Goal} Our goal is to design algorithms to process the data output by the privacy-preserving mechanism, $\releasescoresym$, before its actual release in a manner that improves the privacy-utility tradeoff. Specifically, our goal is to design algorithms that satisfy the four desiderata D1--D4 listed in Section~\ref{sec:intro}.

\section{Theoretical Results}
\label{sec:appendix:theoreticalresults}
We present our main theoretical results in this section.

\subsection{Approach}

We first derive a representation of the set of all possible values in the sorted mean-weight vector $\resultscoresym$ based on the public data. For any paper $\sampleindex \in \paperSet$, we use $\receivescore_{\sampleindex 1}, \receivescore_{\sampleindex 2},\cdots, \receivescore_{\sampleindex \reviewperpaper}$ to denote the $\reviewperpaper$ weights on edges connected to that vertex (paper) in the public data, listed in an arbitrary order. Note that the second subscript of $\receivescore_{\sampleindex j}$ does not correspond to a reviewer identity. The multiset $\{\receivescore_{1 1},\cdots,\receivescore_{\numpapers \reviewperpaper}\}$ is available publicly and for each $\sampleindex \in \paperSet$, the multiset $\{\receivescore_{\sampleindex 1}, \cdots, \receivescore_{\sampleindex \reviewperpaper}\}$ is identical to the multiset $\{\paperreviewerscore_{\sampleindex \sampleindexscd} | \text{reviewer } \sampleindexscd \in \reviewerSet \text{ reviews paper } \sampleindex\}$. Let $\assignmentset$ be a set of weighted bipartite graphs comprising all valid reviewer-paper weights based on the public data, that is, each member of $\assignmentset$ satisfies:
\begin{itemize}
    \item It is a bipartite graph, with the vertices in the two parts corresponding to papers $\paperSet$ and reviewers $\reviewerSet$.
    \item All vertices in $\paperSet$ are $\reviewperpaper$ regular and all vertices in $\reviewerSet$ are $\paperperreviewer$ regular.
    \item The $\reviewperpaper$ edges incident on any vertex $\sampleindex \in \paperSet$ have weights $\receivescore_{\sampleindex 1}$, $\receivescore_{\sampleindex 2}$,\ldots, $\receivescore_{\sampleindex \reviewperpaper}$. 
\end{itemize}

Furthermore, for any graph $\samplegraph \in \assignmentset$, any paper $\sampleindex$ and reviewer $\sampleindexscd$, we define $\paperreviewerscore_{\sampleindex \sampleindexscd}(\samplegraph)$ equal to the weight of edge between $\sampleindex$ and $\sampleindexscd$ if this edge exists, and $\paperreviewerscore_{\sampleindex \sampleindexscd}(\samplegraph) = 0$ otherwise. Then the set $\scoreset$, that comprises all possible values of the sorted mean-weight vector based on public data, is given by
\begin{equation}
    \begin{split}
    \scoreset =& \big\{\resultscoresymproj \in \mathbb{R}^\numreviewers \mid\; \resultscoreproj_1 \le ... \le \resultscoreproj_\numreviewers,~~\exists\samplegraph \in \assignmentset\text{ such that\ }\resultscoreproj_\sampleindexscd = \frac{1}{\paperperreviewer} \sum\limits_{\sampleindex = 1}^\numpapers \paperreviewerscore_{\sampleindex \sampleindexscd}(\samplegraph) \text{ for all }\sampleindexscd \in [\numreviewers]\big\}.
    \end{split}
\end{equation}

Note that the true paper-reviewer graph is also a member of $\assignmentset$ and the true sorted weight vector $\resultscoresym \in \scoreset$. Throughout this section, we consider algorithms that are based on projecting the released data on certain sets. To this end, for any set $\convexSet \subseteq \mathbb{R}^\numreviewers$, we define the projection of vector $\releasescoresym$ on the set $\convexSet$ as
\begin{equation}
     \argmin\limits_{\resultscoresymproj \in \convexSet} \sum\limits_{\sampleindex = 1}^\numreviewers (\resultscoreproj_\sampleindex - \releasescore_\sampleindex)^2.
     \label{EqnClosestPoint}
\end{equation} 

When the privacy-preserving algorithm perturbs the true sorted mean-weight vector $\resultscoresym$, the resulting noisy mean-weight vector $\releasescoresym$ may not lie in the set $\scoreset$. It is thus intuitive to instead replace the resulting vector with the vector in $\scoreset$ closest to it, that is, to instead output the projection~\eqref{EqnClosestPoint} of the vector $\releasescoresym$ with the choice $\convexSet = \scoreset$. 

The following result shows that this intuitive approach can actually \emph{increase} the error. We state and prove this result concretely in the case of additive Laplace noise, but as seen in the proof, the result is much more general.
\begin{proposition}
\label{prop:counter}
There exists a review setting such that the true sorted mean-weight vector $\resultscoresym$, the noisy mean-weight vector $\releasescoresym$ obtained by adding Laplace noise with zero mean and a fixed, non-zero variance to $\resultscoresym$, and the output $\finalscoresym$ of the projection~\eqref{EqnClosestPoint} of $\releasescoresym$ on the set $\convexSet = \scoreset$, are related as
\begin{equation}
    \mathbb{E} \left[ \sum\limits_{\sampleindex = 1}^\numreviewers (\finalscore_\sampleindex - \resultscore_\sampleindex)^2 \right] > \mathbb{E} \left[ \sum\limits_{\sampleindex = 1}^\numreviewers (\releasescore_\sampleindex - \resultscore_\sampleindex)^2\right],
\end{equation}
where the expectation is taken with respect to the noise distribution.
\end{proposition}
The proposition implies that using the closest valid vector violates desideratum~\ref{itmMaintext:desideratum-accuracy} of not reducing the accuracy. The proof of Proposition~\ref{prop:counter} is given in Section~\ref{sec:proof:prop:counter}.

Consequently, in order to ensure desideratum~\ref{itmMaintext:desideratum-accuracy} of not reducing the accuracy is met, we project the noisy data onto a convex set that contains $\scoreset$. The following proposition (proved in Section~\ref{sec:proof:prop:convex}) indicates that such projection can never harm the accuracy, and is a straightforward application of the fact that projection on to convex sets is non-expansive.
\begin{proposition}
\label{prop:convex}
Consider any true sorted mean-weight vector $\resultscoresym$, set $\scoreset$ to comprise all possible true sorted mean-weight vectors, and any arbitrary (noisy mean-weight) vector $\releasescoresym$. Let $\convexSet$ be any closed convex set such that $\scoreset \subseteq \convexSet$. Let $\finalscoresym=(\finalscore_1, ..., \finalscore_\numreviewers)$ be the projection of $\releasescoresym$ on to set $\convexSet$ as in~\eqref{EqnClosestPoint}. Then it must be that
\begin{equation}
    \sum\limits_{\sampleindex = 1}^\numreviewers (\finalscore_\sampleindex - \resultscore_\sampleindex)^2 \le \sum\limits_{\sampleindex = 1}^\numreviewers (\releasescore_\sampleindex - \resultscore_\sampleindex)^2.
\end{equation}
\end{proposition}
Since proposition \ref{prop:convex} holds for all $\releasescoresym$, it follows that if $\releasescoresym$ is obtained by randomly perturbing $\resultscoresym$, then
\begin{equation}
    \mathbb{E} \left[\sum\limits_{\sampleindex = 1}^\numreviewers (\finalscore_\sampleindex - \resultscore_\sampleindex)^2 \right] \le \mathbb{E} \left[\sum\limits_{\sampleindex = 1}^\numreviewers (\releasescore_\sampleindex - \resultscore_\sampleindex)^2 \right].
    \label{eq:propconvex}
\end{equation}
Moreover, for two closed convex sets $\convexSet_1 \subseteq \convexSet_2$, if we have a projection on $\convexSet_2$, then further projecting it on $\convexSet_1$ can never increase the error and can possibly decrease the error.
\textbf{Our goal thus is to project the noisy data on to a (small) convex set that contains all possible true values.}

\subsection{NP-hardness of Projection onto Convex Hull}
The smallest convex set that contains $\scoreset$ is the convex hull of $\scoreset$. Observe that if we could project on to the convex hull, then it can also be used to improve upon the projection on any other convex set. Specifically, if $\finalscoresym$ is the projection of the perturbed data $\releasescoresym$ on some convex set that contains $\scoreset$, and if $\finalscoresym'$ is the projection of $\finalscoresym$ on $\text{convex-hull}(\scoreset)$, then with an argument identical to that in Proposition~\ref{prop:convex} we have that $\sum\limits_{\sampleindex = 1}^\numreviewers (\finalscore'_\sampleindex - \resultscore_\sampleindex)^2 \le \sum\limits_{\sampleindex = 1}^\numreviewers (\releasescore_\sampleindex - \resultscore_\sampleindex)^2 $.

Consequently, in this section we consider the goal of projecting the noisy data onto the convex hull of $\scoreset$. In this case, the final result we would like to output can be represented as choosing $\convexSet = \text{convex-hull}(\scoreset)$ in Equation~\eqref{EqnClosestPoint}. 
Unfortunately, as we show below, projection onto convex-hull($\scoreset$) is NP-hard. 
\begin{theorem}
\label{thm:generalNP}
When $\reviewperpaper = \paperperreviewer > 2$, $\numpapers=\numreviewers$ and $\numreviewers$ is a multiple of $\paperperreviewer$, the problem of projecting noisy data onto convex-hull($\scoreset$) is NP-hard.
\end{theorem}

We prove this result via reducing the $\paperperreviewer$-Partition problem to the projection problem. Given any instance of an $\paperperreviewer$-Partition problem, which is a multiset of integers, we can construct a conference where each paper has a weight from the multiset. We can construct a vector such that the projection result can directly answer the $\paperperreviewer$-Partition problem. The complete proof of Theorem~\ref{thm:generalNP} is provided in Section~\ref{sec:proof:thm:generalNP}.

\begin{remark}
The proof of Theorem~\ref{thm:generalNP} also shows that projection on to $\scoreset$ is NP-hard.
\end{remark}

\subsection{An Efficient Algorithm}
\label{EfficientAlgorithm}
In this section, we present an algorithm that meets the four desiderata D1--D4 listed in Section~\ref{sec:intro}. Since we cannot efficiently project on to the convex hull of $\scoreset$, we must make do with a larger convex set that contains $\scoreset$. We use desideratum~\ref{itmMaintext:desideratum-axiomatic} for guidance on what constitutes a reasonably small set and associated projection.

\subsubsection{Axioms defining desideratum~\ref{itmMaintext:desideratum-axiomatic}}
\label{sec:axioms}
Recall that desideratum~\ref{itmMaintext:desideratum-axiomatic} says that the algorithm should automatically recover the ground truth when the structure of the public data is simple enough. More concretely, we benchmark any algorithm using the following axiomatic properties:
\begin{enumerate}[label=A\arabic*]
    \item \label{itmMaintext:axiomatic-all_same}When all weights are identical, the projection should result in a vector whose entries are all the same as the weight. Formally, if $\receivescore_{\sampleindex \sampleindexscd}=\samplescore\ \forall \sampleindex \in [\numpapers], \sampleindexscd \in [\reviewperpaper]$ for some $\samplescore$, then the output $\finalscoresym$ of the algorithm must be $\finalscore_1=\finalscore_2=\cdots=\finalscore_\numreviewers=\samplescore$.
    \item \label{itmMaintext:axiomatic-ppr1}When $\paperperreviewer = 1$ (that is, each reviewer reviews 1 paper), the projection of any noisy data should result in a sorted vector of all weights. Formally, if $\paperperreviewer = 1$ then the output $\finalscoresym$ of the algorithm must be $(\finalscore_1,\finalscore_2,\cdots,\finalscore_\numreviewers)=\text{sorted}(\receivescore_{1 1},\receivescore_{2 1},\cdots,\receivescore_{\numreviewers 1})$.
    \item \label{itmMaintext:axiomatic-most_zero}When all but one papers have all zero weights, the projection of any noisy data should result in a sorted vector with $(\numreviewers-\reviewperpaper)$ zero entries and the remaining entries equal to $\frac{1}{\paperperreviewer}$ of the weights for the paper that does not have all-zero weights. Formally, if $\receivescore_{\sampleindex \sampleindexscd}=0\ \forall \sampleindex \in \{2,\ldots,\numpapers\}, \sampleindexscd \in [\reviewperpaper]$, then the output $\finalscoresym$ of the algorithm must be $(\finalscore_1,\finalscore_2,\cdots,\finalscore_\numreviewers)=\text{sorted}(\frac{\receivescore_{1 1}}{\paperperreviewer},\frac{\receivescore_{1 2}}{\paperperreviewer},\cdots,\frac{\receivescore_{1 \reviewperpaper}}{\paperperreviewer}, 0, \cdots, 0)$.
\end{enumerate}

\subsubsection{High-level idea behind the algorithm}

The main idea behind our algorithm comprises the following three steps:
\begin{enumerate}[label=\Roman*.]
    \item From the public data, take all tuples of size $\paperperreviewer$ containing weights from different papers into consideration.
    \item Use them to construct lower and upper bounds on every entry of (the unknown vector) $\resultscoresym$.
    \item Project the released data $\releasescoresym$ on the set specified by the aforementioned lower and upper bounds, along with any other problem-specific (convex) constraints.
\end{enumerate}

As one can intuitively see, these three steps imply a projection of the released data on a convex set which includes all valid values of the true data, and hence from Proposition~\ref{prop:convex} it will not reduce the utility (desideratum~\ref{itmMaintext:desideratum-accuracy}). Moreover, the entire algorithm uses only the public data along with the vector $\releasescoresym$ released by the privacy mechanism, and hence does not compromise privacy (desideratum~\ref{itmMaintext:desideratum-privacy}). The idea is general enough to be applied to many forms of the released data, and in what follows, we apply it to release the histogram of the true sorted mean-weight vector. Of course, the devil lies in the details of how these steps are executed, which will determine whether the designed algorithm meets desiderata~\ref{itmMaintext:desideratum-time} and \ref{itmMaintext:desideratum-axiomatic}.

\subsubsection{Full algorithm description}
\label{sec:appendix:fullalgo}
We now describe our algorithm in full detail; we provide an illustrative example subsequently in Section~\ref{sec:appendix:example}. 
Recall that we use $\receivescore_{\sampleindex 1}$, $\receivescore_{\sampleindex 2}, \cdots, \receivescore_{\sampleindex \reviewperpaper}$ to represent the $\reviewperpaper$ weights on edges connected to any vertex (paper) $\sampleindex \in \paperSet$ in the public data. Note that since reviewer identities are not available publicly, the second subscript ``$\sampleindexscd$'' in ``$\receivescore_{\sampleindex \sampleindexscd}$'' has no particular meaning other than capturing the fact that each paper has $\reviewperpaper$ weights.
We use matrix $\scorematrix$ to display all the weights in the public data where
$\scorematrix =
\left[\begin{matrix}
\receivescore_{1 1} & \cdots & \receivescore_{1 \reviewperpaper} \\
\receivescore_{2 1} & \cdots & \receivescore_{2 \reviewperpaper} \\
\vdots & \ddots & \vdots \\
\receivescore_{\numpapers 1} & ... & \receivescore_{\numpapers \reviewperpaper}
\end{matrix}\right].
$

~\\\noindent\textbf{I. Valid weight tuples}

We define a weight tuple as a multiset of $\paperperreviewer$ real values. We say that a tuple is a \emph{valid weight tuple} if it consists of $\paperperreviewer$ weights from distinct papers. In other words, a valid weight tuple contains $\paperperreviewer$ entries of matrix $\scorematrix$ where no two entries are from the same row in $\scorematrix$. We compute $\alltuples$ as the list of all valid weight tuples. In other words, $\alltuples$ contains all the possible weight tuples given by a reviewer. We sort the list $\alltuples$ based on the mean weight of the weight tuples (breaking ties arbitrarily), and henceforth use the notation $\sortedalltuples$ for this sorted list.

~\\\noindent\textbf{II. Lower and upper bounds}
We now compute lower and upper bounds on each entry of $\resultscoresym$ based only on the public data. We create a graph $\tuplegraph$ which has all weight tuples in $\sortedalltuples$ as its vertices. Since each weight tuple in $\sortedalltuples$ corresponds to a vertex in graph $\tuplegraph$, we use the terms ``weight tuples in $\sortedalltuples$'' and ``vertices in $\tuplegraph$'' interchangeably.
There is an edge between two vertices if the two weight tuples do not contain weights from the same entry in $\scorematrix$. Then for each vertex, we define its left chain and right chain as follows. Recall that $\sortedalltuples$ is a sorted list and all of its entries, which are weight tuples, are totally ordered. We use the indices of the tuples (vertices) in this ordering for the following definitions.
\begin{definition}[Left chain]
For any vertex $\tuplevertex$ in $\tuplegraph$, a left chain of $\tuplevertex$ is a simple path in $\tuplegraph$ from $\tuplevertex$ to another vertex such that the indices of the vertices in this path decrease starting from $\tuplevertex$.
\end{definition}
\begin{definition}[Right chain]
For any vertex $\tuplevertex$ in $\tuplegraph$, a right chain of $\tuplevertex$ is a simple path in $\tuplegraph$ from $\tuplevertex$ to another vertex such that the indices of the vertices in this path increase starting from $\tuplevertex$.
\end{definition}
We also define the \emph{length of a chain} to be the number of vertices in the chain. For each vertex, we compute the length of its longest left chain and the length of its longest right chain using dynamic programming.
To compute the length of the longest left chain of a vertex $\tuplevertex$ in $\tuplegraph$, we check the length of the longest left chain of all its neighbors at lower indices. Then the length of the longest left chain of $\tuplevertex$ is the maximum of these neighbors' longest left chain lengths plus one. Similarly, to compute the length of the longest right chain of $\tuplevertex$, we check the length of the longest right chain of all its neighbors at higher indices. The length of the longest right chain of $\tuplevertex$ is the maximum of its neighbors' longest right chain lengths plus one. We store the length of the longest left and right chain of each vertex for subsequent use in the algorithm.

We first present the algorithm to compute a lower bound on $\resultscore_\sampleindex$ for each $\sampleindex \in \reviewerSet$. The computation for upper bounds is analogous to the computation for lower bounds. The algorithm for computing the lower bounds is presented in Algorithm~\ref{alg:lowerbound}.

In more detail, the algorithm uses two criteria to determine if mean of a weight tuple is a lower bound on $\resultscore_\sampleindex$. The criteria are
\begin{enumerate}[label=C\arabic*]
    \item \label{itm:locriterion-chain}The longest left chain of the tuple has length at least $\sampleindex$.
    \item \label{itm:locriterion-uncrossed}In $\scorematrix$, after we mark the $\paperperreviewer$ weights from each tuple considered so far, each row has at most $\numreviewers-\sampleindex$ unmarked entries.
\end{enumerate}
We call the $\numreviewers$ weight tuples that compute $\resultscoresym$ the true weight tuples. The true weight tuple that has mean $\resultscore_\sampleindex$ must have $\sampleindex-1$ weight tuples that have smaller or equal mean to $\resultscore_\sampleindex$. No two reviewers give the same weight so no two true weight tuples contain weights from the same entry in $\scorematrix$. Thus, criterion~\ref{itm:locriterion-chain} is a necessary condition for a weight tuple to be the true weight tuple that computes $\resultscore_\sampleindex$. In addition, there are $\numreviewers-\sampleindex$ entries in $\resultscoresym$ whose values are no smaller than $\resultscore_\sampleindex$. Since no two true weight tuples contain weights from the same paper, each paper can have at most $\numreviewers-\sampleindex$ unused weights for these entries. Therefore, each row of $\scorematrix$ cannot have more than $\numreviewers-\sampleindex$ unmarked entries. Thus, criterion~\ref{itm:locriterion-uncrossed} is necessary for all weights to be assigned among the reviewers. For each entry $\sampleindex \in \reviewerSet$, we choose the valid weight tuple with the smallest mean that satisfies criteria~\ref{itm:locriterion-chain} and~\ref{itm:locriterion-uncrossed} as the lower bound on $\resultscore_\sampleindex$. Hence, it is a valid lower bound.

\begin{algorithm}[tb]
   \caption{Computation of lower bounds}
   \label{alg:lowerbound}
\begin{algorithmic}
   \STATE {\bfseries Input:} matrix $\scorematrix$ of weights, sorted list of weight tuples $\sortedalltuples$
   \STATE {\bf Initialize} $\sampleindex = 1$, set $\paperreviewerscore \in \mathbb{R}^\paperperreviewer$ as the first tuple in $\sortedalltuples$, and all entries of $\scorematrix$ are unmarked.
   \WHILE{$\sampleindex \le \numreviewers$} 
   \STATE{for each weight in the tuple $\paperreviewerscore$, find its corresponding entry in matrix $\scorematrix$ and mark the entry}
   %\IF{$\sampleindex = 1$}
   %\STATE lower bound on $\resultscore_\sampleindex$ = mean of the tuple
   %\STATE $\sampleindex += 1$
   \IF{length of tuple $\paperreviewerscore$'s longest left chain $\ge \sampleindex$ \textbf{and} number of unmarked entries on each row of $\scorematrix \le \numreviewers-\sampleindex$}
   \STATE  lower bound on $\resultscore_\sampleindex$ = mean of all entries of tuple $\paperreviewerscore$
   \STATE $\sampleindex += 1$
   \ENDIF
   \STATE set $\paperreviewerscore$ as the next tuple in $\sortedalltuples$
   \ENDWHILE
\end{algorithmic}
\end{algorithm}

The computation of upper bounds is similar to the above methodology and is presented in Algorithm~\ref{alg:upperbound}. The two criteria we use to determine if mean of a tuple is an upper bound on $\resultscore_\sampleindex$ are
\begin{enumerate}[label=C\arabic*]
\setcounter{enumi}{2}
    \item \label{itm:upcriterion-chain}The longest right chain of the tuple has length at least $\numreviewers-\sampleindex+1$.
    \item \label{itm:upcriterion-uncrossed}In $\scorematrix$, after we mark the $\paperperreviewer$ weights from each tuple considered so far, each row has at most $\sampleindex-1$ unmarked entries.
\end{enumerate}

\begin{algorithm}[tb]
   \caption{Computation of upper bounds}
   \label{alg:upperbound}
\begin{algorithmic}
   \STATE {\bfseries Input:} matrix $\scorematrix$ of weights, sorted list of weight tuples $\sortedalltuples$
   \STATE Initialize $\sampleindex = \numreviewers$, $\paperreviewerscore \in \mathbb{R}^\paperperreviewer$ as the last tuple in $\sortedalltuples$, and all entries of $\scorematrix$ are unmarked.
   \WHILE{$\sampleindex \ge 1$}
   \STATE{for each weight in the tuple $\paperreviewerscore$, find its corresponding entry in matrix $\scorematrix$ and mark the entry}
   \IF{length of tuple $\paperreviewerscore$'s longest right chain $\ge \numreviewers-\sampleindex+1$ \textbf{and} number of unmarked entries on each row of $\scorematrix \le \sampleindex - 1$}
   \STATE upper bound on $\resultscore_\sampleindex$ = mean of all entries of tuple $\paperreviewerscore$
   \STATE $\sampleindex -= 1$
   \ENDIF
   \STATE set $\paperreviewerscore$ as the previous tuple in $\sortedalltuples$
   \ENDWHILE
\end{algorithmic}
\end{algorithm}

~\\\noindent\textbf{III. Projection}
Let $\Lo_\sampleindex$ denote the lower bound we compute on $\resultscore_\sampleindex$ and $\Up_\sampleindex$ denote the upper bound we compute on $\resultscore_\sampleindex$ in part II above. The final output of our algorithm is the solution to the following optimization problem:
\begin{equation}
\begin{split}
    \argmin\limits_{\finalscoresym \in \mathbb{R}^\numreviewers} \sum\limits_{\sampleindex=1}^\numreviewers (\noisescore_\sampleindex - \finalscore_\sampleindex)^2 \text{ such that } \Lo_\sampleindex \le \finalscore_\sampleindex \le \Up_\sampleindex \forall \sampleindex \in \reviewerSet, \\ \sum\limits_{\sampleindex=1}^\numreviewers \finalscore_\sampleindex = \frac{1}{\paperperreviewer} \sum\limits_{\sampleindex=1}^\numpapers \sum\limits_{\sampleindexscd=1}^\reviewperpaper \receivescore_{\sampleindex \sampleindexscd}, \\
    \finalscore_1 \le \finalscore_2 \le \cdots \le \finalscore_\numreviewers.
\end{split}
\end{equation}

This is a convex optimization problem with a quadratic objective and 2$\numreviewers$ linear constraints, and hence is solvable efficiently.

~\\\begin{remark}[Extension to non-uniform paper and reviewer loads]
So far we have assumed that every reviewer reviews the same number of papers and every paper is reviewed by the same number of reviewers. We now discuss how to extend our algorithm to a setting where these assumptions may be violated. First, if the papers are reviewed by different number of reviewers, the exact algorithm as described above continues to hold. Now if different reviewers review different numbers of papers, then we make the following modification to the above algorithm. Let $\paperperreviewerset \subset [\numpapers]$ denote the set of all paper loads on the reviewers, that is, $\paperperreviewer \in \paperperreviewerset \iff $ there is a reviewer who reviews exactly $\paperperreviewer$ papers. Then the set $\alltuples$ computed in the first step of the algorithm includes all weight tuples of size $\paperperreviewer$ for every $\paperperreviewer \in \paperperreviewerset$. The remainder of the algorithm remains identical to that described above. The proof of correctness in these settings follows from the same arguments (given in Section~\ref{sec:proof:thm:correctnessofalgorithm}) as those for the setting of uniform reviewer and paper loads. The algorithm continues to have a computational complexity that is polynomial in $\numreviewers$ and $\numpapers$ (we continue to assume that the maximum number of papers reviewed by any reviewer and the maximum number of reviews by any reviewer are constants).
\end{remark}
%%%%%%%%

\subsubsection{An example}
\label{sec:appendix:example}
In this section, we illustrate our algorithm (described in Section~\ref{sec:appendix:fullalgo}) by means of an illustrative example. 

Consider a case where $\numreviewers = \numpapers = 4$, $\paperperreviewer = \reviewperpaper = 3$. Let 3 papers among the 4 have all 0 weights and the fourth paper has weights 1, 2 and 3. In this example, we can infer that $\resultscorevec = (0, \frac{1}{3}, \frac{2}{3}, 1)$ regardless of assignment. This example reflects axiomatic property~\ref{itmMaintext:axiomatic-most_zero} presented in Section~\ref{EfficientAlgorithm}. We show that our algorithm for computing bounds indeed results in a convex set that contains only the vector $(0, \frac{1}{3}, \frac{2}{3}, 1)$. And thus the projection of any noisy data onto this convex set results in $(0, \frac{1}{3}, \frac{2}{3}, 1)$.

First, we visualize the matrix $\scorematrix$ as
\begin{center}
$
\begin{matrix}
\text{paper 1}: & 0_{1 1} & 0_{1 2} & 0_{1 3} \\
\text{paper 2}: & 0_{2 1} & 0_{2 2} & 0_{2 3} \\
\text{paper 3}: & 0_{3 1} & 0_{3 2} & 0_{3 3} \\
\text{paper 4}: & 1_{4 1} & 2_{4 2} & 3_{4 3}.
\end{matrix} 
$
\end{center}
The subscripts indicate the entries of the weights in $\scorematrix$. Some elements in $\alltuples$ are: $(0_{1 1}, 0_{2 1}, 0_{3 1}),$ $\cdots, (0_{1 3}, 0_{2 3}, 0_{3 3})$, $(0_{1 1}, 0_{1 2}, 1_{4 1})$, $\cdots, (0_{1 3}, 0_{2 3}, 1_{4 1})$, $(0_{1 1}, 0_{1 2}, 2_{4 2}), \cdots, (0_{1 1}, 0_{1 2}, 3_{4 3})$. After we sort $\alltuples$ based on the mean of the weight tuples, we get $\sortedalltuples$ where the first 27 tuples have mean 0, followed by 27 tuples with mean $\frac{1}{3}$, 27 tuples with mean $\frac{2}{3}$, and 27 tuples with mean 1. We construct graph $\tuplegraph$ in which for instance, there is an edge between the tuples $(0_{1 1}, 0_{2 1}, 0_{3 1})$ and $(0_{1 2}, 0_{2 2}, 0_{3 2})$ since all six weights in these two tuples correspond to different entries in $\scorematrix$. On the other hand, there is no edge between the tuples $(0_{1 1}, 0_{2 1}, 1_{4 1})$ and $(0_{1 2}, 0_{2 2}, 1_{4 1})$ because they both contain weight $1_{4 1}$.

The first tuple in $\sortedalltuples$ meets both the criteria so lower bound on $\resultscore_1$ is mean of the first tuple, which is a tuple with mean 0. Therefore, lower bound on $\resultscore_1$ is 0. Without loss of generality, the first tuple is $(0_{1 1}, 0_{2 1}, 0_{3 1})$ and we mark the corresponding entries in $\scorematrix$. Now the matrix $\scorematrix$ can be visualized as follows (where we mark any entry when we need to):
\begin{center}
$
\begin{matrix}
\text{paper 1}: & \cancel{0_{1 1}} & 0_{1 2} & 0_{1 3} \\
\text{paper 2}: & \cancel{0_{2 1}} & 0_{2 2} & 0_{2 3} \\
\text{paper 3}: & \cancel{0_{3 1}} & 0_{3 2} & 0_{3 3} \\
\text{paper 4}: & 1_{4 1} & 2_{4 2} & 3_{4 3}.
\end{matrix}
$
\end{center}
To compute a lower bound on $\resultscore_2$, we start from the second tuple in $\sortedalltuples$. Since there are 27 tuples that have mean zero, the second tuple still has mean zero. However, we do not choose any tuple that has mean zero due to criterion~\ref{itm:locriterion-uncrossed} from the algorithm. Choosing any $(0,0,0)$ tuple leaves all 3 entries of row 4 in $\scorematrix$ unmarked, and thus will not leave row 4 with at most 2 unmarked entries. Therefore, we will only stop at the first tuple that has mean $\frac{1}{3}$. Without loss of generality, we choose tuple $(0_{1 1}, 0_{2 1}, 1_{4 1})$ and mark the corresponding entries in $\scorematrix$. This will leave the matrix $\scorematrix$ as
\begin{center}
$
\begin{matrix}
\text{paper 1}: & \cancel{0_{1 1}} & \cancel{0_{1 2}} & \cancel{0_{1 3}} \\
\text{paper 2}: & \cancel{0_{2 1}} & \cancel{0_{2 2}} & \cancel{0_{2 3}} \\
\text{paper 3}: & \cancel{0_{3 1}} & \cancel{0_{3 2}} & \cancel{0_{3 3}} \\
\text{paper 4}: & \cancel{1_{4 1}} & 2_{4 2} & 3_{4 3}.
\end{matrix}
$
\end{center}
For a similar reason, we do not choose any tuple that has mean $\frac{1}{3}$ to be a lower bound on $\resultscore_3$ as it would not leave row 4 of $\scorematrix$ with at most 1 unmarked entry. So we choose the first tuple that has mean $\frac{2}{3}$ and a lower bound on $\resultscore_3$ is $\frac{2}{3}$. Lastly, a lower bound on $\resultscore_4$ is computed using the first tuple that has mean 1.

Now we can look at the computation of upper bounds using the proposed algorithm. Upper bound on $\resultscore_4$ is taken as mean of the last tuple, which is a tuple with mean 1. Therefore, an upper bound on $\resultscore_4$ is 1. In addition, we mark two entries of weight 0 and one entry of weight 3. Without loss of generality, we mark entries $0_{1 1}, 0_{2 1}, 3_{4 3}$. Now the matrix $\scorematrix$ can be visualized as 
\begin{center}
$
\begin{matrix}
\text{paper 1}: & \cancel{0_{1 1}} & 0_{1 2} & 0_{1 3} \\
\text{paper 2}: & \cancel{0_{2 1}} & 0_{2 2} & 0_{2 3} \\
\text{paper 3}: & 0_{3 1} & 0_{3 2} & 0_{3 3} \\
\text{paper 4}: & 1_{4 1} & 2_{4 2} & \cancel{3_{4 3}}.
\end{matrix} 
$
\end{center}
To compute an upper bound on $\resultscore_3$, we start from the second to last tuple in $\sortedalltuples$. Since there are 27 tuples that have mean 1, the second to last tuple still has mean 1. However, we do not choose any tuple with mean 1 due to criterion~\ref{itm:upcriterion-chain} from the algorithm. Any $(0,0,3)$ tuple does not have a right chain of length at most 2 because all tuples with value $(0,0,3)$ are not connected due to the uniqueness of the weight 3. Therefore, we will only stop at the first tuple that has mean $\frac{2}{3}$ as it has a right chain of length 2. Since we have encountered all combinations of $(0,0,3)$, the matrix $\scorematrix$ after we choose a tuple $(0,0,2)$ becomes \begin{center}
$
\begin{matrix}
\text{paper 1}: & \cancel{0_{1 1}} & \cancel{0_{1 2}} & \cancel{0_{1 3}} \\
\text{paper 2}: & \cancel{0_{2 1}} & \cancel{0_{2 2}} & \cancel{0_{2 3}} \\
\text{paper 3}: & \cancel{0_{3 1}} & \cancel{0_{3 2}} & \cancel{0_{3 3}} \\
\text{paper 4}: & 1_{4 1} & \cancel{2_{4 2}} & \cancel{3_{4 3}}
\end{matrix}.
$
\end{center}
For a similar reason, we do not choose any tuple that has mean $\frac{2}{3}$ to be an upper bound on $\resultscore_2$ as it would not have a right chain of length at least 3. So we choose the first tuple we encounter that has mean $\frac{1}{3}$ and an upper bound on $\resultscore_2$ is $\frac{1}{3}$. Lastly, an upper bound on $\resultscore_1$ is computed using the first tuple that has mean 0.

Thus, the bounds on $\resultscoresym = \resultscorevec$ are $0 \le \resultscore_1 \le 0$, $\frac{1}{3} \le \resultscore_2 \le 
\frac{1}{3}$, $\frac{2}{3} \le \resultscore_3 \le \frac{2}{3}$ and $1 \le \resultscore_4 \le 1$. Along with the conditions that $\resultscore_1 + \resultscore_2 + \resultscore_3 + \resultscore_4 = 2$ and $\resultscore_1 \le \resultscore_2 \le \resultscore_3 \le \resultscore_4$, the only possible value of $\resultscoresym$ is $(0, \frac{1}{3}, \frac{2}{3}, 1)$. Thus the output of our algorithm is the singleton set $\{(0, \frac{1}{3}, \frac{2}{3}, 1)\}$. Projection of any data to the convex set $\{(0, \frac{1}{3}, \frac{2}{3}, 1)\}$ results in $(0, \frac{1}{3}, \frac{2}{3}, 1)$, which is consistent with axiomatic property~\ref{itmMaintext:axiomatic-most_zero}.

\subsubsection{Guarantees of Our Algorithm}
In this section, we evaluate our algorithm with respect to the four desiderata listed in Section~\ref{sec:intro}. We first prove the correctness of the algorithm in terms of projection on to an appropriate set. 
\begin{theorem}
\label{thm:correctnessofalgorithm} The algorithm projects noisy data onto a convex set that contains all true values.
\end{theorem}
The proof of this theorem is given in Section~\ref{sec:proof:thm:correctnessofalgorithm}. This result, combined with Theorem~\ref{prop:convex}, guarantees that our algorithm does not increase the error. Thus, our algorithm satisfies desideratum~\ref{itmMaintext:desideratum-accuracy}. In addition, since our algorithm uses only the public data for post processing, it satisfies desideratum~\ref{itmMaintext:desideratum-privacy}.

We now discuss the computational complexity of our algorithm. In our setting, we assume $\paperperreviewer$ and $\reviewperpaper$ to be constants. Since $\numreviewers \cdot \paperperreviewer = \numpapers \cdot \reviewperpaper$, the number of papers, $\numpapers$, is polynomial in the number of reviewers, $\numreviewers$.
\begin{theorem}
\label{thm:efficiencyofalgorithm}
The algorithm has polynomial time complexity in the number of reviewers.
\end{theorem}
So our algorithm satisfies desideratum~\ref{itmMaintext:desideratum-time}. While the algorithm is polynomial time in $\numreviewers$ and $\numpapers$, it is exponential in $\paperperreviewer$. In practice $\paperperreviewer$ is usually a constant, and is frequently small~\cite{shah2018design}. The proof of this theorem is given in Section~\ref{sec:proof:thm:efficiencyofalgorithm}.

We finally visit desideratum~\ref{itmMaintext:desideratum-axiomatic} -- of being able to return an exact answer when it can easily be deduced from public data.
\begin{theorem}
\label{thm:axiomatic}
The algorithm satisfies the axiomatic properties A1, A2 and A3 defined in Section~\ref{sec:axioms}.
\end{theorem}
We have thus shown that our proposed algorithm meets all four desiderata D1--D4.

%%%%%%%%%%%%%%%%%%%%%%%%%

\section{Simulations}
\label{sec:experiment}
In this section, we conduct synthetic simulations to evaluate the performance of our algorithm. We synthetically generate a conference review setting in one of several ways as described below. In each of the settings, the number of reviewers equals the number of papers, and each reviewer reviews 2 papers and each paper is reviewed by two reviewers. The assignment of reviewers to papers is done uniformly at random subject to given load constraints. The weight given by any reviewer to any reviewed paper is drawn from a beta distribution. For preserving privacy, we consider the common method of adding i.i.d.\ Laplace noise (with mean zero and variance 2) to each component of the true sorted mean-weight vector.

We evaluate the following three methods of releasing the sorted mean-weight vector, which includes our proposed algorithm and two baselines:
\begin{itemize}
    %\item {\bf Noiseless} where no noise is added (and hence no privacy is guaranteed);
    \item {\bf Noisy} where Laplace noise is added but no post-processing is performed;
    \item {\bf Baseline projection} where the noisy data is post-processed via projecting onto a convex set which constrains the sum of all entries, the value of each entry in terms of the range of weights (0 to 1), and imposes a monotonicity constraint; We project on the set $\{\finalscoresym \in \mathbb{R}^\numreviewers | 0 \le \finalscore_\sampleindex \le 1 \forall \sampleindex \in \reviewerSet, \ \sum\limits_{\sampleindex=1}^\numreviewers \finalscore_\sampleindex = \frac{1}{\paperperreviewer} \sum\limits_{\sampleindex=1}^\numpapers \sum\limits_{\sampleindexscd=1}^\reviewperpaper \receivescore_{\sampleindex \sampleindexscd}, \ \finalscore_1 \le \finalscore_2 \le \cdots \le \finalscore_\numreviewers\}$.
    \item {\bf Our algorithm} where the noisy data is post-processed via our algorithm described in Section~\ref{EfficientAlgorithm}.
\end{itemize}
The code for our algorithm is available at \url{https://github.com/wenxind/privacy-utility-tradeoff-in-peer-review-data}.

\begin{figure}[!b]
\centering
\begin{subfigure}{.33\textwidth}	
\includegraphics[width=\textwidth]{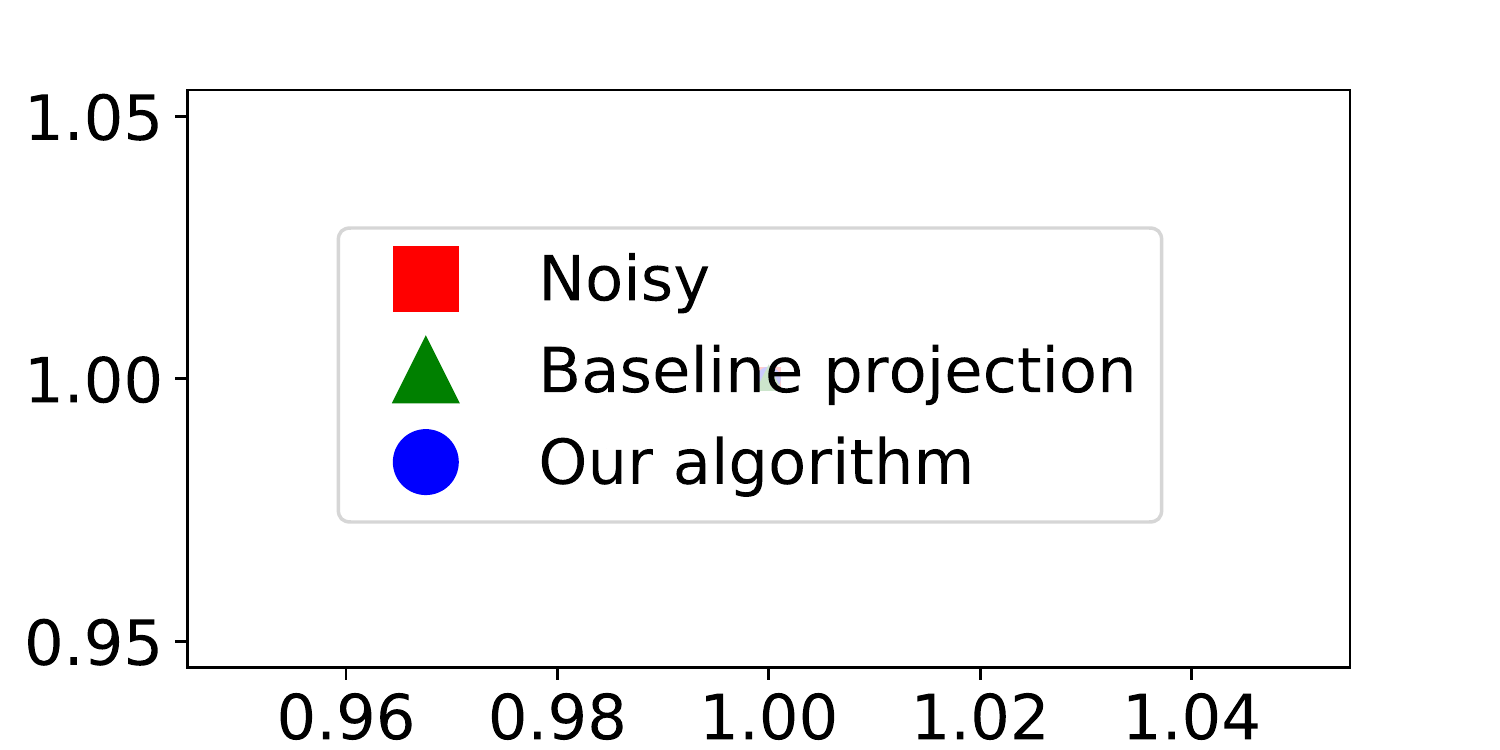}
%\caption{Beta(2,2)\label{FigBeta(2,2)}}
\end{subfigure}\hfill
\begin{subfigure}{.33\textwidth}	
\includegraphics[width =\textwidth]{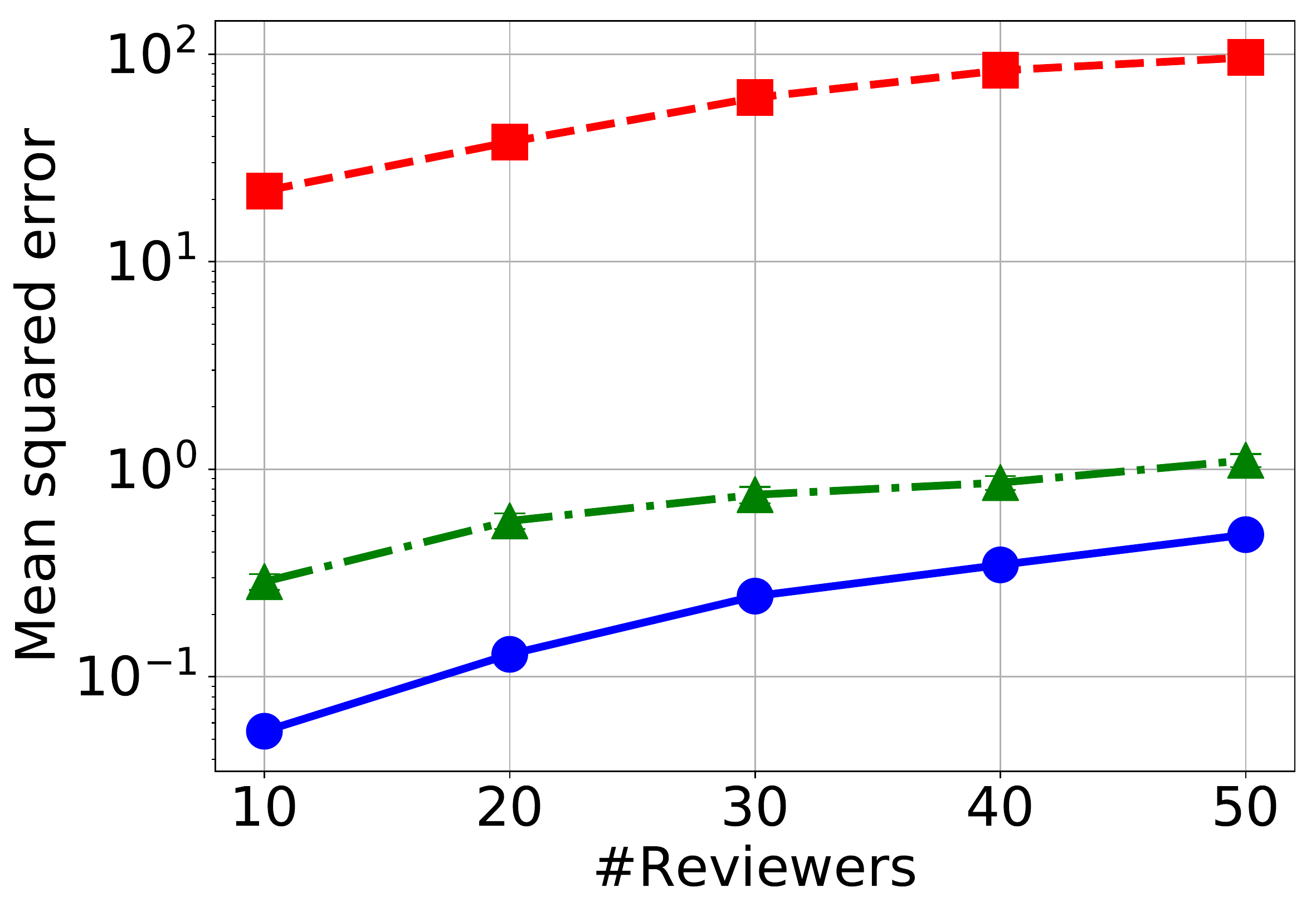}
\caption{Beta(5,1)\label{FigBeta(5,1)}}
\end{subfigure}\hfill
\begin{subfigure}{.33\textwidth}	
\includegraphics[width =\textwidth]{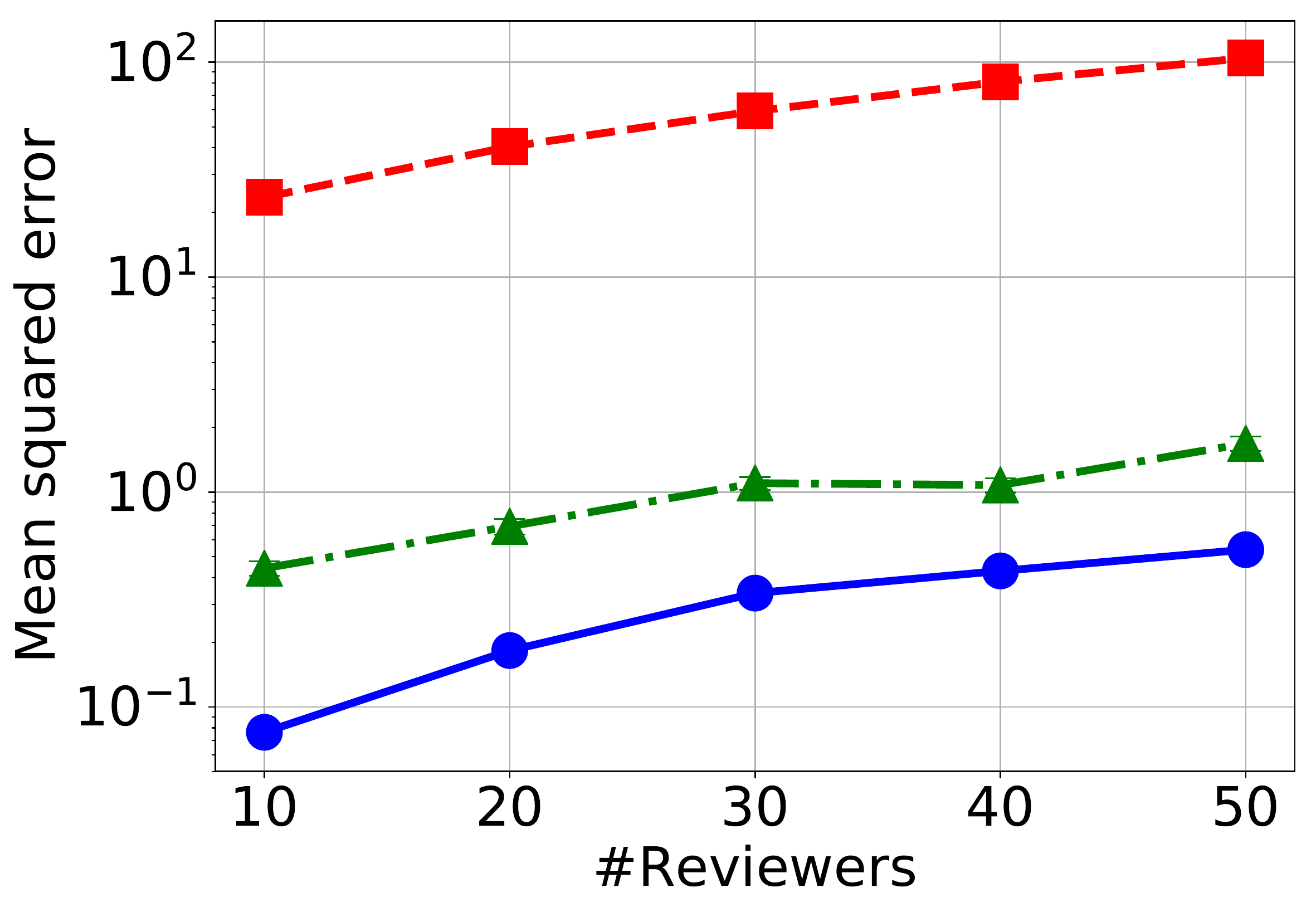}
\caption{Beta(2,5)\label{FigBeta(2,5)}}
\end{subfigure}\hfill
\vspace{.5cm}
\begin{subfigure}{.33\textwidth}	
\includegraphics[width =\textwidth]{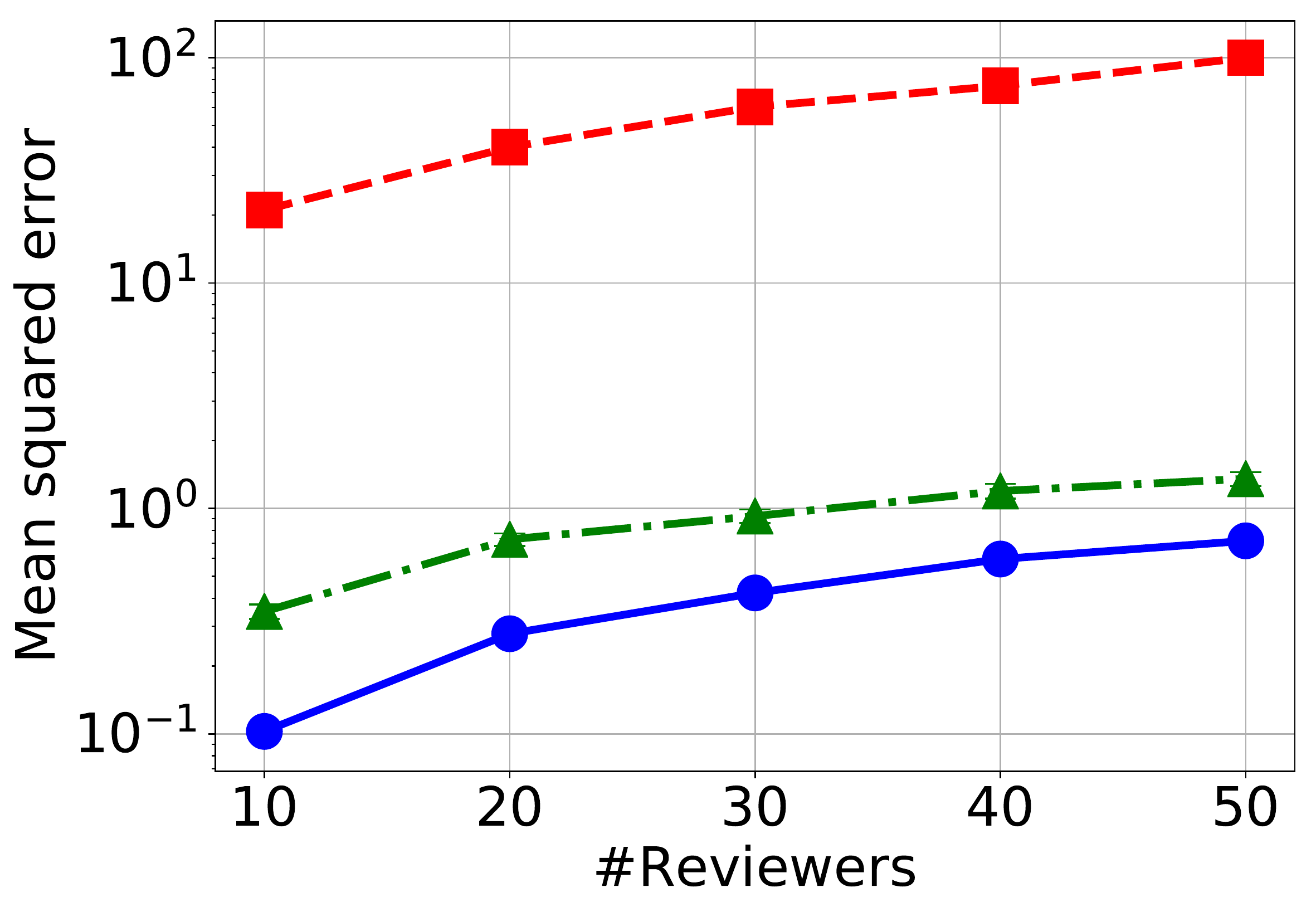}
\caption{Beta(1,3)\label{FigBeta(1,3)}}
\end{subfigure}\hfill
\begin{subfigure}{.33\textwidth}	
\includegraphics[width =\textwidth]{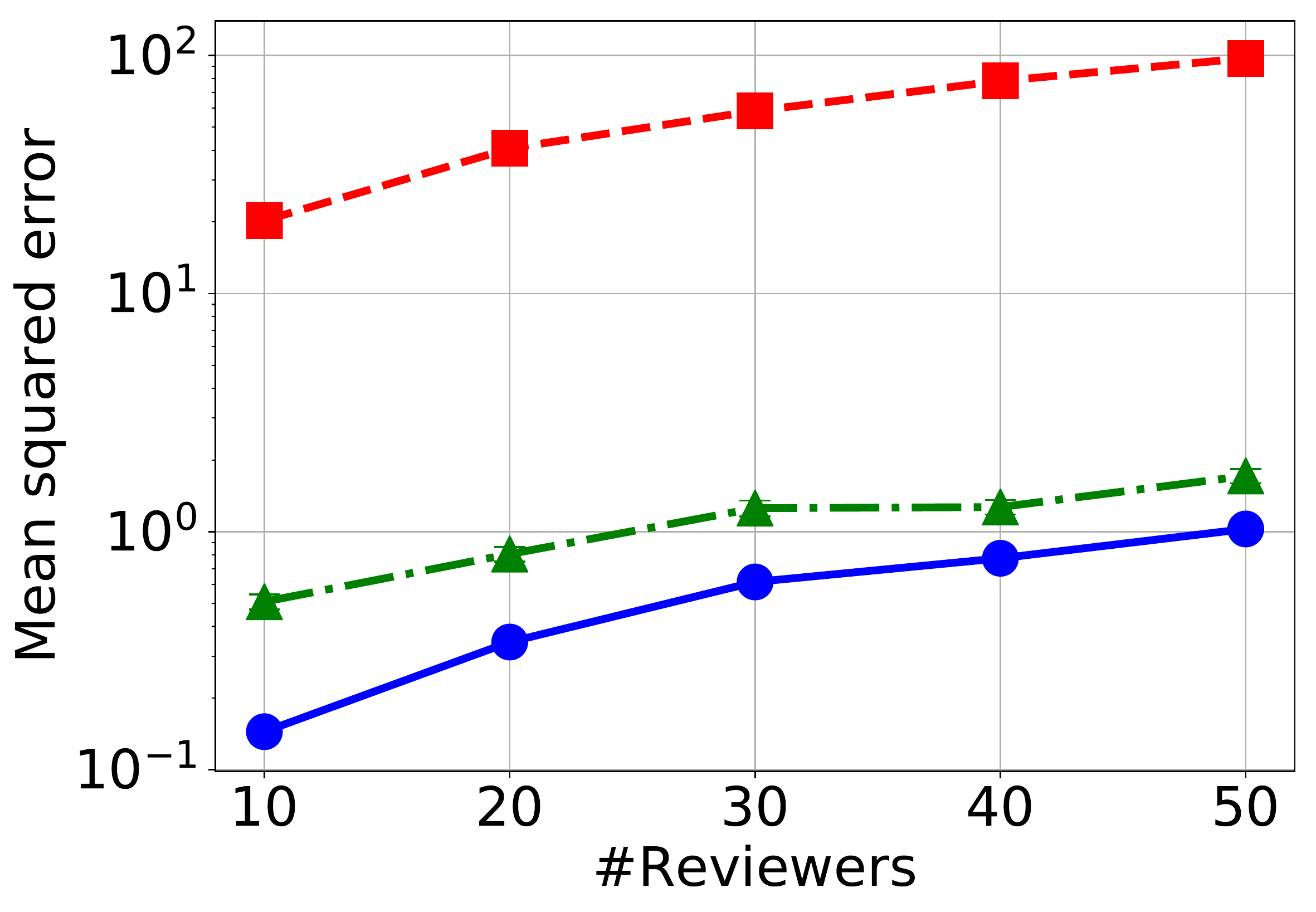}
\caption{Beta(2,2)\label{FigBeta(2,2)}}
\end{subfigure}\hfill
\begin{subfigure}{.33\textwidth}	
\includegraphics[width =\textwidth]{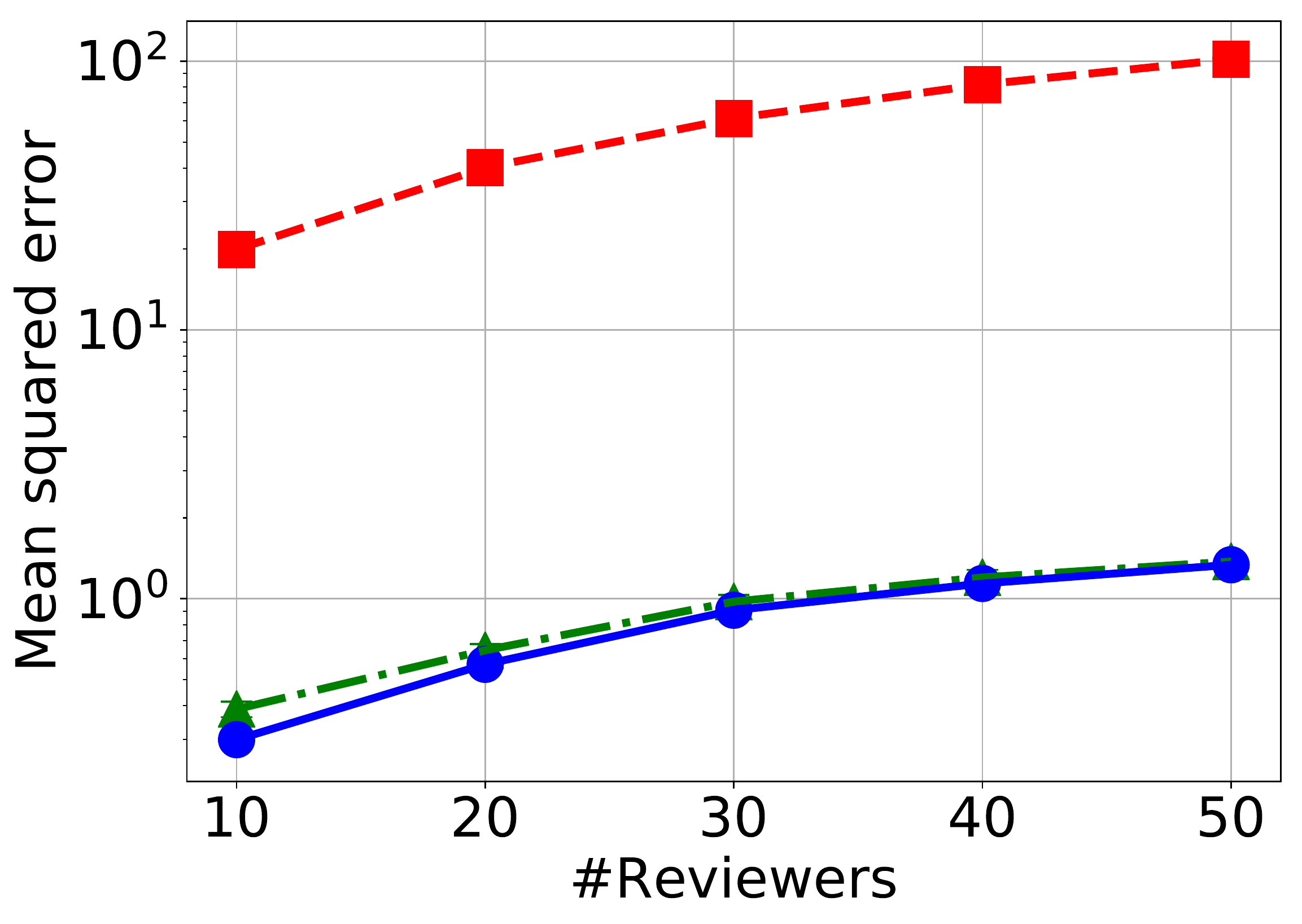}
\caption{Beta(0.5,0.5)\label{FigBeta(0.5,0.5)}}
\end{subfigure}\hfill
\vspace{.5cm}
\begin{subfigure}{.33\textwidth}	
\includegraphics[width =\textwidth]{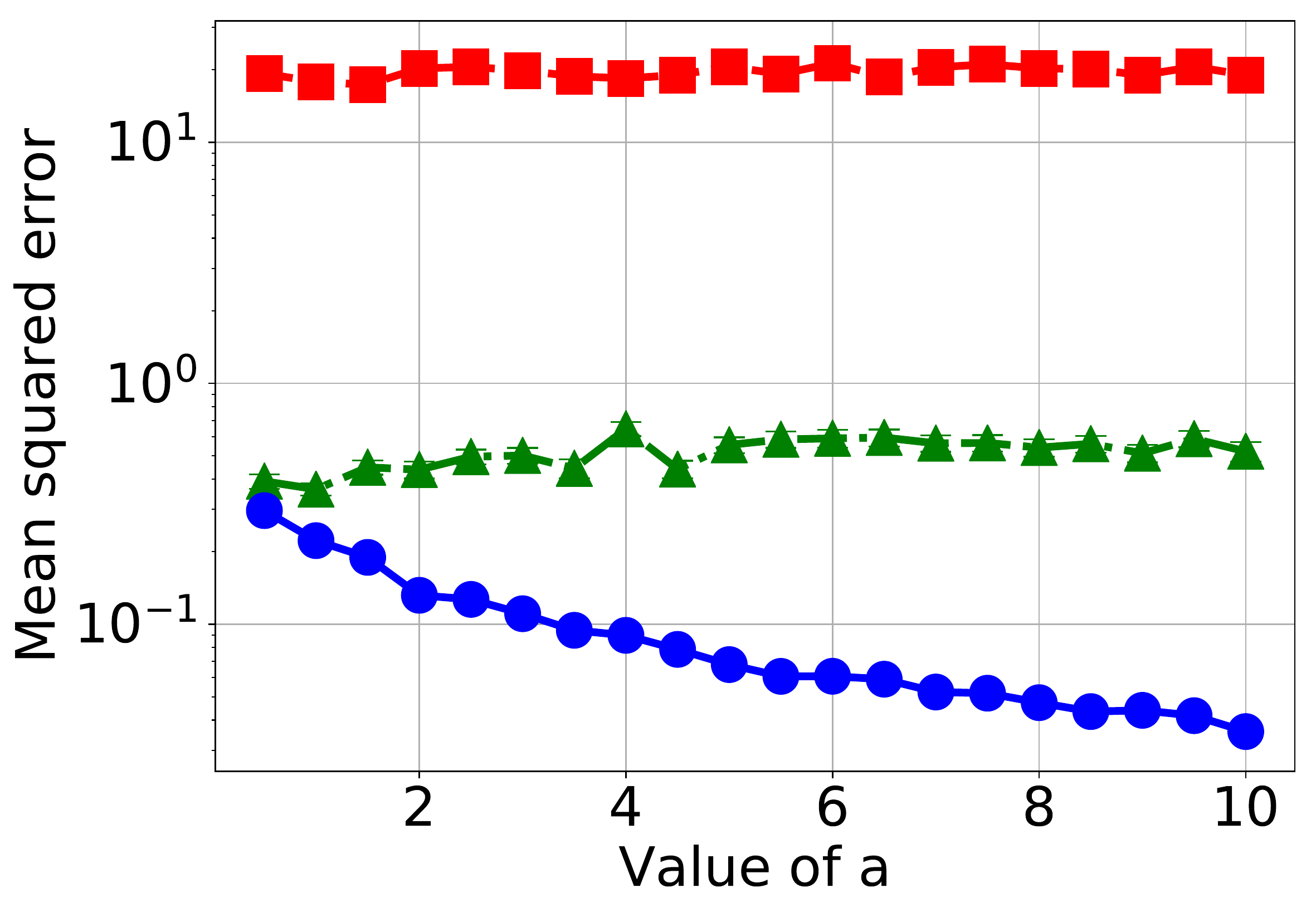}
\caption{Beta(a, a) with 10 reviewers\label{FigBeta(a, a)}}
\end{subfigure}\hfill
\begin{subfigure}{.33\textwidth}	
\includegraphics[width =\textwidth]{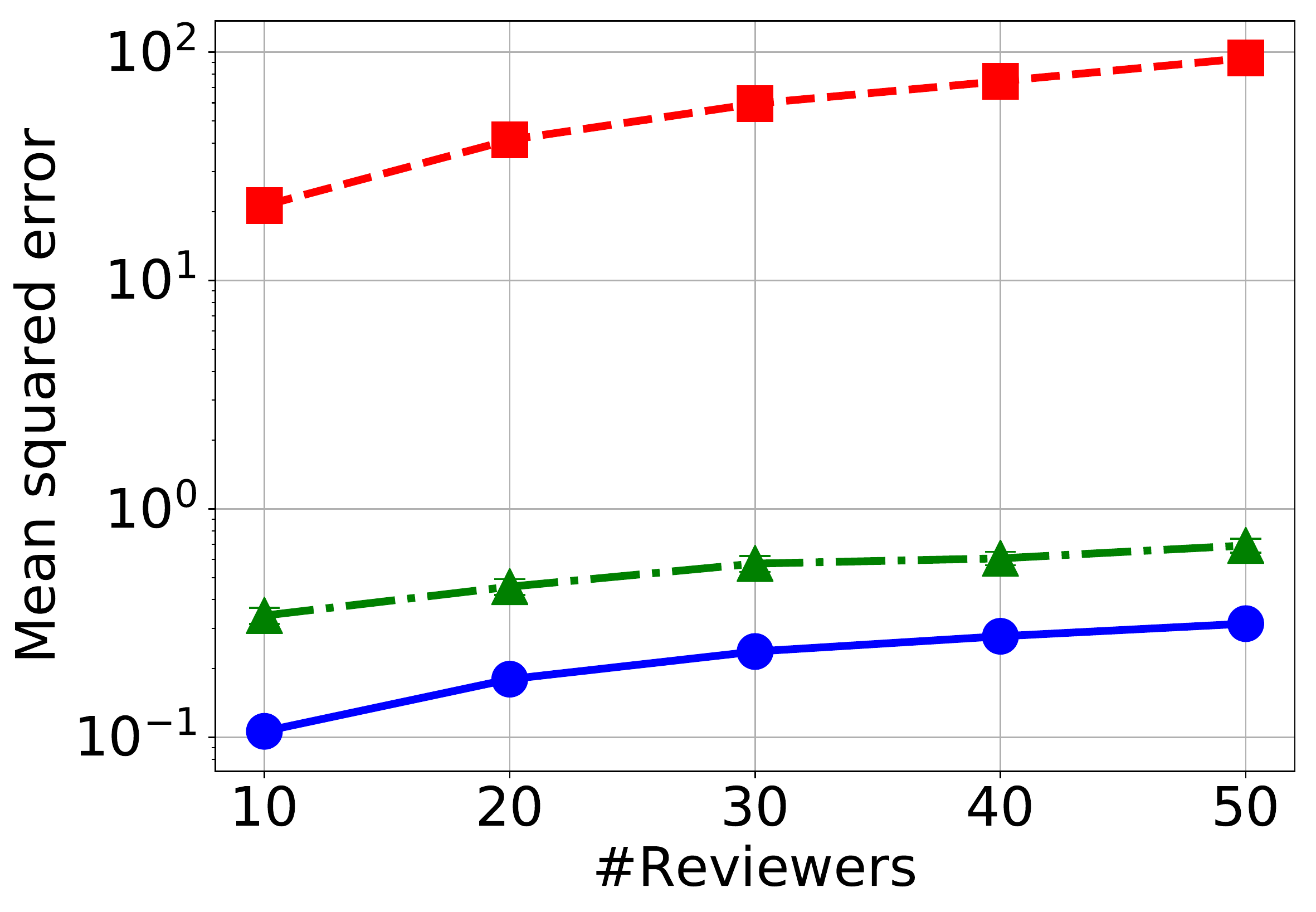}
\caption{Beta(a, b) with a $\in$ [2] and b $\in$ [\numreviewers]\label{FigBeta(1-2, 1-n)}}
\end{subfigure}\hfill
\begin{subfigure}{.33\textwidth}	
\includegraphics[width =\textwidth]{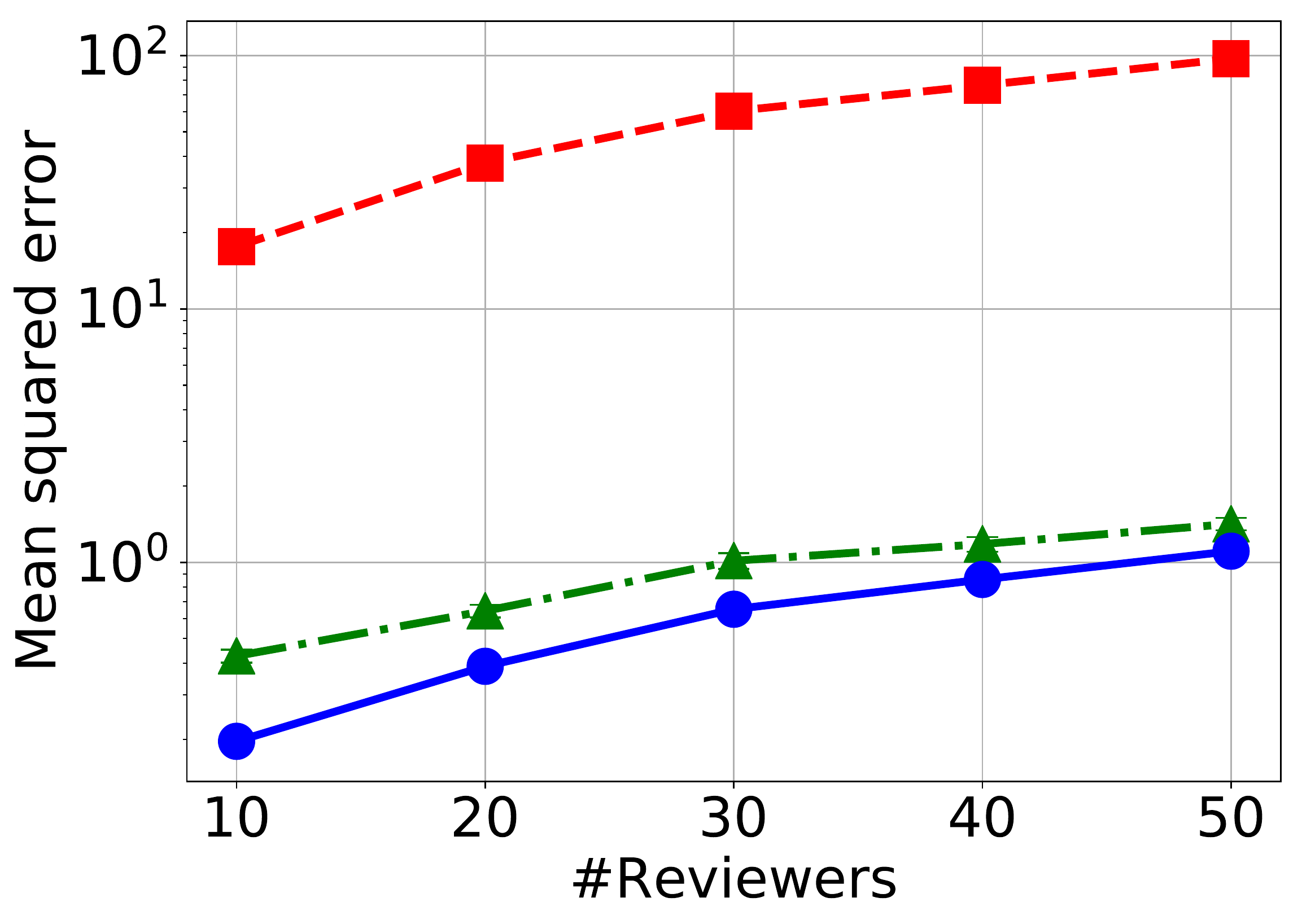}
\caption{Beta(a, b) with a, b $\in$ [\numreviewers] \label{FigBeta(i, j)}}
\end{subfigure}\hfill
\caption{Simulation results\label{FigExperiments}. The y-axes of all plots are on a logarithmic scale.}
\end{figure}

The simulations compute the mean squared error between the true sorted mean-weight vector $\resultscoresym$ and the output from each of these three methods, that is, $\sum_{\sampleindex=1}^{\numreviewers} ( \finalscore_\sampleindex - \resultscore_\sampleindex)^2$ where $\finalscoresym$ is the output of any of these algorithms. Note that in the figures, the error bars (standard error of the mean) are plotted but not visible in most cases since they are too small.

We now describe the method for generating the weights in each simulation, and refer the reader to the corresponding plots. Note that the y-axes (representing the mean squared error) on each of the plots is on a logarithmic scale. 
\begin{itemize}
    \item In Figure~\ref{FigBeta(5,1)}---\ref{FigBeta(0.5,0.5)}, the number of reviewers ranges from 10 to 50. The weights are all i.i.d.\ and are generated from the beta distribution specified in the corresponding subcaption. 
    
    \item In Figure~\ref{FigBeta(a, a)}, the number of reviewers is fixed at 10. On the x-axis, we vary a parameter $a \in \{0.5,1,\ldots,10\}$. For each value of $a$, we draw all weights i.i.d.\ from the beta$(a, a)$ distribution. 
    
    \item In Figure~\ref{FigBeta(1-2, 1-n)}, we again vary the number of reviewers $\numreviewers$ on the x-axis. For any paper $\sampleindex \in \reviewerSet$, one weight is generated from beta$(1, \sampleindex)$ and the other weight is generated from beta$(2, \sampleindex)$ independently.
    
    \item In Figure~\ref{FigBeta(i, j)}, whenever any paper $\sampleindex \in \paperSet$ is reviewed by reviewer $\sampleindexscd \in \reviewerSet$, the weight of that review is generated from beta$(\sampleindex, \sampleindexscd)$.
\end{itemize}
All in all, these simulations reveal that our algorithm can lead to a multi-fold improvement in the utility (accuracy) while not compromising the privacy.

\section{Discussion and Future Work}

We take the first steps towards designing methods for privacy-preserving release of peer-review data, and posit the need for much more research on this topic to address the important challenge of improving peer review. There are several open problems of interest. First, we propose an algorithm to improve the privacy-utility tradeoff, and even though it has a polynomial time complexity in the number of reviewers, its computation time is practically infeasible for large conferences such as NeurIPS, ICML or AAAI which have 5000-10000 papers and several thousand reviewers. A useful direction of future work is to improve the computational complexity of the algorithm to make it operate in ``practical time''. Second, we currently consider releasing histograms of mean scores given by each reviewer, and it is of considerable theoretical and practical interest to enable privacy-preserving release of other peer-review data, such as properties of the reviewer graph, reviewer bids, and other functions of the scores. Third, in this work we separate out the privacy component of data release from the post-processing component which improves utility. It is of interest to jointly design both components which may yield an even better tradeoff. Finally, it is of interest to design methods that can utilize data from multiple conferences, while preserving the privacy in each conference, for improving the peer-review process in any subsequent conference.

\clearpage
\section*{Acknowledgments}
This work was supported in part by NSF CAREER award 1942124.

\bibliographystyle{unsrt}
\bibliography{reference,refs-weina}

\appendix

\section{Appendix: Proofs}
In the appendix, we present complete proofs of the results claimed in the main text.

\subsection{Proof of Proposition~\ref{prop:counter}}\label{sec:proof:prop:counter}
We will prove the proposition using a counter example. Assume the true value $\resultscoresym = 0$ and the set of all possible values $\scoreset = \{-4, -2, 0, 2, 4\}$. The noisy data $\releasescoresym = \resultscoresym + \noisescoresym$ where $\noisescoresym$ is a Laplace random variable with probability density function $\noisescore(x) = 0.5 e^{-|x|}$.

Without projection, the expected error incurred by the noise is $\int_{-\infty}^\infty 0.5 e^{-|x|} x^2 d x = 2$. But if we project the noisy data on the set $\scoreset$ and get result $\finalscoresym$, the expected error after the projection is computed as $16 \int_{-\infty}^{-3} 0.5 e^{-|x|} d x + 4 \int_{-3}^{-1} 0.5 e^{-|x|} d x + 4 \int_{1}^{3} 0.5 e^{-|x|} d x + 16 \int_{3}^{\infty} 0.5 e^{-|x|} d x = 2.06896$, which is greater than the expected error without projection. Thus, projecting on the set that contains all true values could decrease the accuracy of data.

\subsection{Proof of Proposition~\ref{prop:convex}}\label{sec:proof:prop:convex}
It is known that projection on a closed convex set is non-expansive~\cite{bauschke2011convex}. Since $\resultscoresym$ results from a valid assignment, it is contained in $\scoreset$. Therefore it is contained in any closed convex set that contains $\scoreset$. Projection of $\releasescoresym$ onto any such convex set will not increase its squared error from $\resultscoresym$. Therefore, proposition~\ref{prop:convex} holds.

\subsection{Proof of Theorem \ref{thm:generalNP}}\label{sec:proof:thm:generalNP}

We will prove the NP-hardness by reducing the $\paperperreviewer$-Partition problem, which is NP-hard~\cite{babel1998thek}, to the problem of projecting noisy data onto convex hull of $\scoreset$. The $\paperperreviewer$-Partition problem where $\paperperreviewer > 2$ is defined as follows.

\begin{definition}
$\paperperreviewer$-Partition problem: Given a multi-set $\samplelist = \{\samplelistitem_1, \samplelistitem_2, ..., \samplelistitem_\samplelistsize\}$ of $\samplelistsize$ non-negative integers where $\samplelistsize$ is a multiple of $\paperperreviewer$, decide if we can partition $\samplelist$ into $\frac{\samplelistsize}{\paperperreviewer}$ subsets such that each subset has size $\paperperreviewer$ and the sums of all subsets are the same.
\end{definition}

Consider any instance of the $\paperperreviewer$-Partition problem with $\samplelist = \{\samplelistitem_1, \samplelistitem_2, ..., \samplelistitem_\numreviewers\}$, where $\samplelistitem_\sampleindex \ge 0$ and $\numreviewers$ is a multiple of $\paperperreviewer$. 
Now we construct a peer-review dataset where there are $\numreviewers$ reviewers and $\numreviewers$ papers, each reviewer reviews $\paperperreviewer$ papers and each paper receives $\paperperreviewer$ reviews. Note that the number of reviewers is the same as the number of elements in $\samplelist$. Let each paper $\sampleindex$ has weight $\samplelistitem_\sampleindex$ and $\paperperreviewer-1$ zero weights, and $\samplelistavg$ denote the average of all elements in $\samplelist$, i.e.,
\begin{equation}
    \samplelistavg = \frac{1}{\numreviewers} \sum\limits_{\sampleindex=1}^\numreviewers \samplelistitem_\sampleindex.
\end{equation}
Let $\noisydata = (0, ..., 0, \samplelistavg, ..., \samplelistavg)$ be a vector of $\numreviewers$ entries whose last $\frac{\numreviewers}{\paperperreviewer}$ entries all have value $\samplelistavg$. Let $\partitionset = \samplelist \cup \{0, ..., 0\}$ be the multiset containing all values of $\samplelist$ and $\numreviewers \cdot (\paperperreviewer-1)$ zeros. Then the projection problem is to project $\noisydata$ onto the convex hull of $\scoreset$ defined for this peer-review dataset.

The reduction from the $\paperperreviewer$-Partition problem to the projection problem constructed above is as follows.
If the solution to the projection problem is $\noisydata$ itself, we return True for the $\paperperreviewer$-Partition problem; otherwise we return False.

We first prove the correctness of the reduction. Suppose $\samplelist$ can be $\paperperreviewer$-partitioned into $\frac{\numreviewers}{\paperperreviewer}$ subsets of equal sums. Then we can partition $\partitionset$ into subsets of size $\paperperreviewer$ where these subsets are the subsets that give the $\paperperreviewer$-partition of $\samplelist$ and subsets that consist of $\paperperreviewer$ zeros. By Lemma~\ref{partitionlemma} below, this partition of $\partitionset$ gives a valid assignment for the peer-review problem, and thus $\noisydata = (0, ..., 0, \samplelistavg, ..., \samplelistavg)$ corresponds to a valid assignment. Therefore, the projection of $\noisydata$ is itself.
The proof of Lemma~\ref{partitionlemma} is presented in Section~\ref{sec:proof:partitionlemma}.

\begin{lemma}
\label{partitionlemma}
In the setting described above, any $\paperperreviewer$-partition of the $\numreviewers \cdot \paperperreviewer$ weights in $\partitionset$, i.e., any partition of $\partitionset$ into subsets of size $\paperperreviewer$, can be interpreted as a valid assignment such that subset $\sampleindex$ corresponds to the weights from reviewer $\sampleindex$ given to $\paperperreviewer$ distinct papers.
\end{lemma}

Next, suppose that the projection of $\noisydata$ is itself. We show that $\samplelist$ can be $\paperperreviewer$-partitioned into subsets of equal sums. We first claim that $\noisydata$ must correspond to a valid assignment itself. To see this, suppose $\noisydata=(0, ..., 0, \samplelistavg, ..., \samplelistavg)$ is a convex combination of some sorted mean weight vectors. Then these vectors must all have value $\samplelistavg$ for their last $\frac{\numreviewers}{\paperperreviewer}$ entries since each of these mean weight vector is sorted. Due to the sum constraint, these mean weight vectors have to be $(0, ..., 0, \samplelistavg, ..., \samplelistavg)$. Next we note that in the assignment given by $\noisydata$, each reviewer who has an average weight of $\samplelistavg$ must give $\paperperreviewer$ weights with values from $\samplelist$ due to the pigeonhole principle. Therefore, the $\frac{\numreviewers}{\paperperreviewer}$ subsets each of which consists of weights given by one of the last $\frac{\numreviewers}{\paperperreviewer}$ reviewers form an $\paperperreviewer$-partition of $\samplelist$ with equal sums.

Finally, we prove the efficiency of the reduction. Since the construction of $\noisydata$ has $\mathcal{O}(\numreviewers)$ time complexity and the construction of $\partitionset$ has $\mathcal{O}(1)$ time complexity, the reduction has $\mathcal{O}(\numreviewers)$ time complexity, which is polynomial in the size of the input. Thus, the reduction can be done efficiently, which completes the proof.

\subsection{Proof of Lemma~\ref{partitionlemma}}\label{sec:proof:partitionlemma}
Fix $\paperperreviewer$, we will prove the lemma by induction on $\numreviewers$, the number of reviewers, which is the same as the number of papers.

Base case: when $\numreviewers = \paperperreviewer$, every reviewer reviews all papers, so any $\paperperreviewer$ partition of the weights can be validly assigned to reviewers.

Inductive hypothesis: suppose when there are fewer than $\numreviewers$ reviewers for an $\numreviewers > \paperperreviewer$, every $\paperperreviewer$ partition of the weights forms a valid assignment for reviewers.

Consider when there are $\numreviewers$ reviewers and $\numreviewers$ papers. Without loss of generality, assume all weights in $\samplelist$ are non-zero. Consider an $\paperperreviewer$ partition of the set $\partitionset$. We will argue in two cases based on whether there is a subset that contains exactly one non-zero weight. 
\begin{enumerate}
\item Case 1: In the partition, if there is a subset with exactly one non-zero weight.

Without loss of generality, assume that the subset with exactly one non-zero weight contains $\resultscorefirst$ and the subset is $\{\resultscorefirst, 0, ..., 0\}$. We denote the subset $\firstpartitionset$. In $\firstpartitionset$, there are $\paperperreviewer - 1$ zero weights and a non-zero weight $\resultscorefirst$ from $\samplelist$. Since paper 1 receives $\paperperreviewer$ weights in total, we can remove $\firstpartitionset$, paper 1 and reviewer 1.

Now we are left with $\numreviewers-1$ reviewers and papers. The removal does not affect the number of reviews received by the rest of the papers. We still have each paper getting $\paperperreviewer$ weights. Among the weights, there is one non-zero weight from $\samplelist$ and $\paperperreviewer-1$ zero weights. By the inductive hypothesis, the rest of the subsets in the partition form a valid assignment of $\partitionset \setminus \firstpartitionset$. We can assign the weights to $\numreviewers-1$ reviewers. 

We then add $\firstpartitionset$ back to the assignment. Since reviewer 1 $\paperperreviewer$ weights to paper 1, it is not valid. We can solve this by swapping the zero weights in $\firstpartitionset$ with zeros in other subsets. We need to make $\paperperreviewer-1$ swaps. We label the rest of the subsets $\restpartitionset_2, \ldots, \restpartitionset_\numreviewers$ where $\restpartitionset_2$ is the subset that contains most non-zero weights and the labels go in decreasing order based on the number of non-zero weights contained in a subset. We look at the rest of the subsets in the order of their labels.

Since none of the rest of the subsets contain any weight from paper 1, swapping a zero weight from paper 1 into any of these subsets will nor affect the validity of the subset. There are at least $\numreviewers-1 - \frac{\numreviewers-1}{\paperperreviewer}$ subsets that contain at least a zero weight. Since $\numreviewers > \paperperreviewer$ and $\paperperreviewer > 2$, $\numreviewers-1 - \frac{\numreviewers-1}{\paperperreviewer} = \frac{(\paperperreviewer-1)(\numreviewers-1)}{\paperperreviewer} \ge \paperperreviewer-1$. Thus, we have enough subsets to swap the zero weights from paper 1 in.

Then we make sure the zero weights swapped into $\firstpartitionset$ will not come from the same paper. We label the zeros in $\firstpartitionset$ with index $1, \ldots, \paperperreviewer-1$. Suppose there are no subsets that do not contain any zero weights. Then when we need to swap out the zero weight at index $\sampleindex$ in $\firstpartitionset$, there are at most $\numreviewers - \sampleindex$ non-zero weights in the untouched subsets due to the order we look at the subsets. There are $\numreviewers - \sampleindex$ untouched subsets as well. Then there exists an untouched subset that contains $\sampleindex$ zero weights. Since at this stage $\firstpartitionset$ has already completed $\sampleindex-1$ swaps, we can find a zero weight from the untouched subset to swap so that the zero weight does not come from the same paper as the zero weights from previous swaps. Note that if we have any subset that does not contain any zero weight or we skip some subsets due to conflict of papers, then the fraction of non-zero weights left and untouched subsets will be even smaller. So we are guaranteed to find a proper zero weight to swap. Thus, we can make $\paperperreviewer-1$ swaps of the zero weights to $\firstpartitionset$ and make all subsets valid assignments of weights. Such swaps do not affect the values in each subset.

Therefore, such partition can result in a valid assignment of the $\numreviewers \cdot \paperperreviewer$ scores among $\numreviewers$ reviewers.

\item Case 2: In the partition, if there are no subsets with exactly one non-zero weight.

Since $\samplelist$ contains $\numreviewers$ elements and there are $\numreviewers$ subsets, by pigeon hole principle, there must be a subset $\firstpartitionset$ that contains all zero weights.

Without loss of generality, we find the subset that contains $\resultscorefirst$ and then swap $\resultscorefirst$ with a zero weight in $\firstpartitionset$. This results in $\firstpartitionset' = \{\resultscorefirst, 0, \ldots, 0\}$.

Now we have a subset that contains exactly 1 weight from $\samplelist$. Like in case 1, we remove the subset, reviewer 1 and paper 1. We can find a valid assignment of the rest of the weights to $\numreviewers-1$ reviewers. Then we will put $\firstpartitionset'$ back to the assignment. Currently all weights in $\firstpartitionset'$ are from paper 1. We identify the subset where $\resultscorefirst$ comes from, and swap $\resultscorefirst$ back into the subset with a zero weight there. Since the subset can not contain any weights from paper 1, we can safely put $\resultscorefirst$ back without having two weights from the same paper.

After the swap, $\firstpartitionset'$ has $\paperperreviewer-1$ zero weights from paper 1 and a zero weight from a different paper, say paper 2. We need to make $\paperperreviewer-2$ swaps for the zeros in $\firstpartitionset'$. We label the rest of the subsets $\restpartitionset_2, \ldots, \restpartitionset_\numreviewers$ where $\restpartitionset_2$ is the subset that contains most non-zero weights and the labels go in decreasing order based on the number of non-zero weights contained in a subset. We look at the rest of the subsets in the order of their labels.

Since none of the rest of the subsets contain any weight from paper 1, swapping a zero weight from paper 1 into any of these subsets will nor affect the validity of the subset. In the worst case, there exists a subset that contains $\resultscorefirst$ and there are at most $\frac{\numreviewers-2}{\paperperreviewer-1}$ subsets that only contains a zero weight from paper 2 because such tuples cannot contain $\resultscoresecond$. Then there are at least $\numreviewers-1 - \frac{\numreviewers-2}{\paperperreviewer-1} -1$ subsets that we can swap the zero weights in $\firstpartitionset'$. Since $\numreviewers > \paperperreviewer$ and $\paperperreviewer > 2$, $\numreviewers-1 - \frac{\numreviewers-2}{\paperperreviewer-1} -1 = \frac{(\paperperreviewer-2)(\numreviewers-2)}{\paperperreviewer-1} \ge \paperperreviewer-2$. Thus, we have enough subsets to swap the zero weights from paper 1 in.

We keep a zero weight from paper 1 in $\firstpartitionset$ and label the rest of the zero weights in $\firstpartitionset$ with index $1, \ldots, \paperperreviewer-2$. Suppose there are no subsets that do not contain any zero weights. Then when we need to swap the zero weight at index $\sampleindex$ in $\firstpartitionset$, there are at most $\numreviewers - \sampleindex$ non-zero weights in the untouched subsets due to the order we look at the subsets. There are $\numreviewers - \sampleindex$ untouched subsets as well. Then there exists a subset that contains $\sampleindex+1$ zero weights. Since at this stage $\firstpartitionset$ has already completed $\sampleindex-1$ swaps, we can find a zero weight to swap that does not conflict with the weights from previous swaps and not from paper 2 either. Note that if we have any subset that does not contain any zero weight or we skip some subsets due to conflict of papers, then the fraction of non-zero weights left and untouched subsets will be even smaller. So we are guaranteed to find a proper zero weight to swap. Thus, we can make $\paperperreviewer - 2$ swaps of the zero weights to $\firstpartitionset'$ and makes all subsets valid assignments of weights. Such swaps do not affect the value in each subset.

Therefore, such partition can result in a valid assignment of the $\numreviewers \cdot \paperperreviewer$ weights among $\numreviewers$ reviewers.
\end{enumerate}
In conclusion, any $\paperperreviewer$-partition of $\partitionset$ can be interpreted as a valid assignments of weights to $\numreviewers$ reviewers.

\subsection{Proof of Theorem~\ref{thm:correctnessofalgorithm}}\label{sec:proof:thm:correctnessofalgorithm}
We would like to show that the convex set contains $\scoreset$. We will show that the bounds are indeed lower and upper bounds on each entry.

We will first show that the lower bounds computed by the algorithm are correct. 

Assume for the sake of contradiction, there exists an assignment such that $\resultscore_\sampleindex$ is less than the lower bound on $\resultscore_\sampleindex$ we computed, denoted as $\resultscoreproj_\sampleindex$. We use $\tuplevertex$ to denote the tuple that results in $\resultscore_\sampleindex$ and use $\tuplevertex'$ to denote the tuple that we choose in the algorithm that has mean $\resultscoreproj_\sampleindex$. Since $\tuplevertex$ is a valid assignment, it is the sum of $\paperperreviewer$ weights from $\paperperreviewer$ distinct papers. Since $\sortedalltuples$ contains all such tuples, it contains $\tuplevertex$. And since $\resultscore_\sampleindex < \resultscoreproj_\sampleindex$, we encountered $\tuplevertex$ before we encounter $\tuplevertex'$ in $\sortedalltuples$. We did not choose $\tuplevertex$ as the tuple for lower bound due to its violation of either criterion~\ref{itm:locriterion-chain} or criterion~\ref{itm:locriterion-uncrossed}.

If $\tuplevertex$ violates criterion~\ref{itm:locriterion-chain}, it does not have a left chain of size at least $\sampleindex$. There cannot be $\sampleindex-1$ weight tuples each containing $\paperperreviewer$ weights from different papers such that they all have mean no larger than $\resultscore_\sampleindex$. Otherwise they form a left chain of length $\sampleindex$. So $\tuplevertex$ cannot have its mean appear at entry $\sampleindex$ in $\resultscoresym$.

If $\tuplevertex$ violates criterion~\ref{itm:locriterion-uncrossed}, there exists a row that has more than $\numreviewers-\sampleindex$ unmarked entries in $\scorematrix$. The weights of the unmarked entries have not been encountered so far, which indicates that any tuple that contains the weights from unmarked entries has mean no less than $\resultscore_\sampleindex$. Otherwise, we would have encountered the weight before $\tuplevertex$ and mark its entry. We know that there are $\numreviewers-\sampleindex$ reviewers who has mean weight no less than $\resultscore_\sampleindex$. In addition, there are more than $\numreviewers-\sampleindex$ weights left for at least one paper. By Pigeon Hole Principle, there exists a reviewer gives a weight tuple that contains two weights from the same paper. However, no two weights from the same paper can be in the same tuple since one reviewer cannot give 2 weights to the same paper. So $\tuplevertex$ cannot have its mean appear at entry $\sampleindex$ in $\resultscoresym$.

Thus, $\resultscore_\sampleindex$ cannot be a value for entry $\sampleindex$ in $\resultscoresym$. The value $\resultscoreproj_\sampleindex$ we computed is indeed a lower bound on that entry.

Following a similar argument, we can prove the correctness of the upper bounds from the algorithm.

\subsection{Proof of Theorem~\ref{thm:efficiencyofalgorithm}}\label{sec:proof:thm:efficiencyofalgorithm}
We will show that the proposed algorithm has polynomial time complexity in the number of reviewers. There are $\numreviewers \cdot \paperperreviewer$ weights, so the size of $\alltuples$, denoted $|\alltuples|$, has size at most $\binom{\numreviewers \cdot \paperperreviewer}{\paperperreviewer}$, which is of complexity $\mathcal{O}(\numreviewers^\paperperreviewer)$. Sorting $\alltuples$ has $\mathcal{O}(|\alltuples| \log (|\alltuples|))$ time complexity, which is still polynomial in $\numreviewers$. There are $\binom{|\alltuples|}{2}$ pairs of vertices to examine for edges. Therefore, constructing $\tuplegraph$ is of polynomial time in $\numreviewers$. To compute the length longest left chain and right chain of a vertex, we can make use of a dynamic programming algorithm that only requires us to loop through $\sortedalltuples$ once to compute length of longest left chain of all vertices and loop one more time to compute the length of longest right chain. For each vertex, we examine at most all its neighbors, which is of size polynomial in $\numreviewers$. Lastly, after all preparation work, for each vertex, we take $\mathcal{O}(1)$ time to check criteria~\ref{itm:locriterion-chain} and and~\ref{itm:upcriterion-chain} at most $\mathcal{O}(\numpapers)$ time to check criteria~\ref{itm:locriterion-uncrossed} and~\ref{itm:upcriterion-uncrossed}. Since $\numpapers \le \numreviewers \cdot \paperperreviewer$, both operations are polynomial in $\numreviewers$. Thus, the proposed algorithm computes the bounds in time polynomial in $\numreviewers$.

We will use quadratic programming to project noisy data onto the convex set and there are 2$\numreviewers$ linear constraints. This operation is also polynomial in $\numreviewers$.

Thus, the proposed algorithm has time complexity that is polynomial in $\numreviewers$.

\subsection{Proof of Theorem~\ref{thm:axiomatic}}\label{sec:proof:thm:axiomatic}
Axiomatic property~\ref{itmMaintext:axiomatic-all_same}: When all weights are the same, all weight tuples have the same mean, which equals the weight. Thus, all lower and upper bounds have the same value as the weight. The convex set contains a single vector and projection of any noisy data on such convex set will result in the vector, whose entries are all the same as the weight. 

Axiomatic property~\ref{itmMaintext:axiomatic-ppr1}: When $\paperperreviewer = 1$, there are exactly $\numreviewers$ weight tuples, each containing one weight. We will choose the same weight tuple for lower bound and upper bound on $\resultscore_\sampleindex$. The mean of the chosen weight tuple is the weight of rank $\sampleindex$ among all $\numreviewers$ weights. Therefore, the convex set contains exactly one vector, which is the sorted vector of all weights. Projection of any noisy data onto this convex set will result in the vector of sorted weights.

Axiomatic property~\ref{itmMaintext:axiomatic-most_zero}: When all except for one paper receives all zero weights, computation of lower bound on $\resultscore_\sampleindex$ when $\sampleindex < \numreviewers - \reviewperpaper$ will choose a tuple whose weights are all zeros. When $\sampleindex \ge \numreviewers - \reviewperpaper$, computation of lower bound will choose a tuple that contains a nonzero weights due to criterion~\ref{itm:locriterion-uncrossed}. Similarly, to compute an upper bound on $\resultscore_\sampleindex$ when $\sampleindex \ge \numreviewers - \reviewperpaper$, we will choose a tuple with a nonzero weight due to the criterion~\ref{itm:upcriterion-chain}. But when $\sampleindex < \numreviewers - \reviewperpaper$, the algorithm will choose a tuple with all zero weights. The example we present in Section~\ref{sec:appendix:example} illustrates this process. Therefore, the convex set again contains only a vector who has $\numreviewers-\reviewperpaper$ zero entries. Projection of any noisy data will result in this vector.

\end{document}